\documentclass[twocolumn, prc, amssymb, superscriptaddress,aps,preprintnumbers,amsmath,floatfix]{revtex4}
\usepackage{multirow}
\usepackage{graphicx}% Include figure files
\usepackage{dcolumn}% Align table columns on decimal point
\usepackage{bm}% bold math
\usepackage{color}
\usepackage{amsmath}%[fleqn]
\usepackage{wasysym}
\usepackage{float}
\usepackage{array,url,kantlipsum}
\usepackage{verbatim}
\usepackage{xcolor}
\usepackage{siunitx}
\usepackage{CJK}
\usepackage{dsfont}
\usepackage{textcomp}
\newcolumntype{P}[1]{>{\centering\arraybackslash}p{#1}}
\usepackage{array}
\usepackage{upgreek}

\newcommand{\ket}[1]{\left| #1 \right\rangle}
\newcommand{\bra}[1]{\left\langle #1 \right|}

\begin{document}

%%%%%%%%%%%%%%%%%%%%%%%%%%%%%%%%%%%%%%%%%%%%%%%%%%%%%%%%%%%%%%%%%%%%%%%%%%%%%%%%%
\title{Nuclear coherent population transfer to  the  $^{229m}$Th isomer using x-ray pulses}

%%%%%%%%%%%%%%%%%%%%%%%%%%%%%%%%%%%%%%%%%%%%%%%%%%%%%%%%%%%%%%%%%%%%%%%%%%%%%%%%%

\author{Tobias \surname{Kirschbaum}}
\email{tobias.kirschbaum@fau.de}
\affiliation{Department of Physics, Friedrich-Alexander-Universit\"at Erlangen-N\"urnberg, 91058 Erlangen, Germany}

\author{Nikolay \surname{Minkov}}
%\email{nminkov@inrne.bas.bg}
\affiliation{Institute of Nuclear Research and
Nuclear Energy, Bulgarian Academy of Sciences, Tzarigrad Road 72, BG-1784
Sofia, Bulgaria}
\affiliation{Max-Planck-Institut f\"ur Kernphysik, Saupfercheckweg 1, 69117 Heidelberg, Germany}

\author{Adriana P\'alffy}
\email{adriana.palffy-buss@fau.de}
%\affiliation{ Max-Planck-Institut f\"ur Kernphysik, Saupfercheckweg 1, 69117 Heidelberg, Germany}
\affiliation{Department of Physics, Friedrich-Alexander-Universit\"at Erlangen-N\"urnberg, 91058 Erlangen, Germany}
\affiliation{Max-Planck-Institut f\"ur Kernphysik, Saupfercheckweg 1, 69117 Heidelberg, Germany}
%%%%%%%%%%%%%%%%%%%%%%%%%%%%%%%%%%%%%%%%%%%%%%%%%%%%%%%%%%%%%%%%%%%%%%%%%%%%%%%%%

\date{\today}

\begin{abstract}
Population of the 8 eV $^{229m}$Th isomer via the second nuclear excited state at 29.19 keV by means of coherent x-ray pulses is investigated theoretically. We focus on two nuclear coherent population transfer schemes  using partially overlapping x-ray pulses known from quantum optics: stimulated Raman adiabatic passage (STIRAP), and successive $\pi$ pulses. Numerical results are presented for three possible experimental setups. Our results identify the Gamma Factory as the most promising scenario, where two ultraviolet pulses combined with relativistically accelerated ions deliver the required intensities for efficient isomer population. Our simulations require knowledge of the in-band and cross-band  nuclear transition probabilities. We give theoretically predicted values for the latter and discuss them in the context of recent experiments. 

\end{abstract}
%%%%%%%%%%%%%%%%%%%%%%%%%%%%%%%%%%%%%%%%%%%%%%%%%%%%%%%%%%%%%%%%%%%%%%%%%%%%%%%%%

\maketitle

\section{Introduction \label{intro}}
Among the entire nuclear chart, the  $^{229}$Th isotope presents a unique isomer which is the lowest known metastable first excited state at energy $\approx \SI{8}{\electronvolt}$ \cite{Wense_Nature_2016,Beck_78eV_2007_corrected, Sikorsky2020}.
This very low energy is an exception among otherwise typically much higher nuclear level energies, and has the peculiarity that it could be in principle addressed by a narrow-band vacuum ultraviolet (VUV) laser. 
This renders $^{229}$Th an ideal candidate for a nuclear clock exceeding the accuracy of present atomic clocks \cite{Peik_Clock_2003, Campbell_Clock_2012, Peik_Clock_2015,Peik_2021}. In addition to a very precise and novel nuclear frequency standard, a nuclear clock would have significant impact also for other applications, for instance  the detection of dark matter \cite{Peik_2021,fadeev20}, investigating the temporal variation of fundamental constants \cite{Flambaum06,Flambaum2020}, the construction of the first nuclear laser \cite{Tkalya_NuclLaser_2011}, or improving the global positioning system \cite{thirolf19}.

A major hurdle towards the nuclear clock experimental implementation is the  relatively large uncertainty on the isomeric state energy. %Zitat
Recently, the energy of the isomeric state was reported to be $E_{\text{iso}}= \SI{8.28+-0.17}{\electronvolt}$ \cite{Seiferle_EnTh229m_2019} using the direct measurement of internal conversion electrons, $E_{\text{iso}}= \SI{8.30+-0.92}{\electronvolt}$\cite{Yamaguchi_EnTh229m_2019} from the determination of transition rates and energies from the second excited state at \SI{29.19}{\kilo \electronvolt} or $E_{\text{iso}}= \SI{8.10+-0.17}{\electronvolt}$ \cite{Sikorsky2020} from $\gamma$-spectroscopic measurements of the $\alpha$-decay of the parent nucleus $^{233}$U. Up to this point, direct laser excitation failed, and the radiative decay of the isomer could not be observed. The isomer energy measurements are based on two indirect excitation methods. The first method populates the isomeric state with a probability of $\approx \SI{2}{\percent}$ during the $\alpha$-decay of $^{233}$U \cite{Thielking2018}. However, the nuclear decay is a purely statistical phenomenon and therefore the population transfer is not controllable. 
The second  method incoherently populates the isomeric level via x-ray pumping of the second excited state at \SI{29.19}{\kilo \electronvolt}. In the measurements in Ref.~\cite{masuda}, this state is populated by synchrotron radiation pulses and partially decays to the 8 eV isomeric state. 

\begin{figure}[!htp]
    \centering
\includegraphics[width = 8cm]{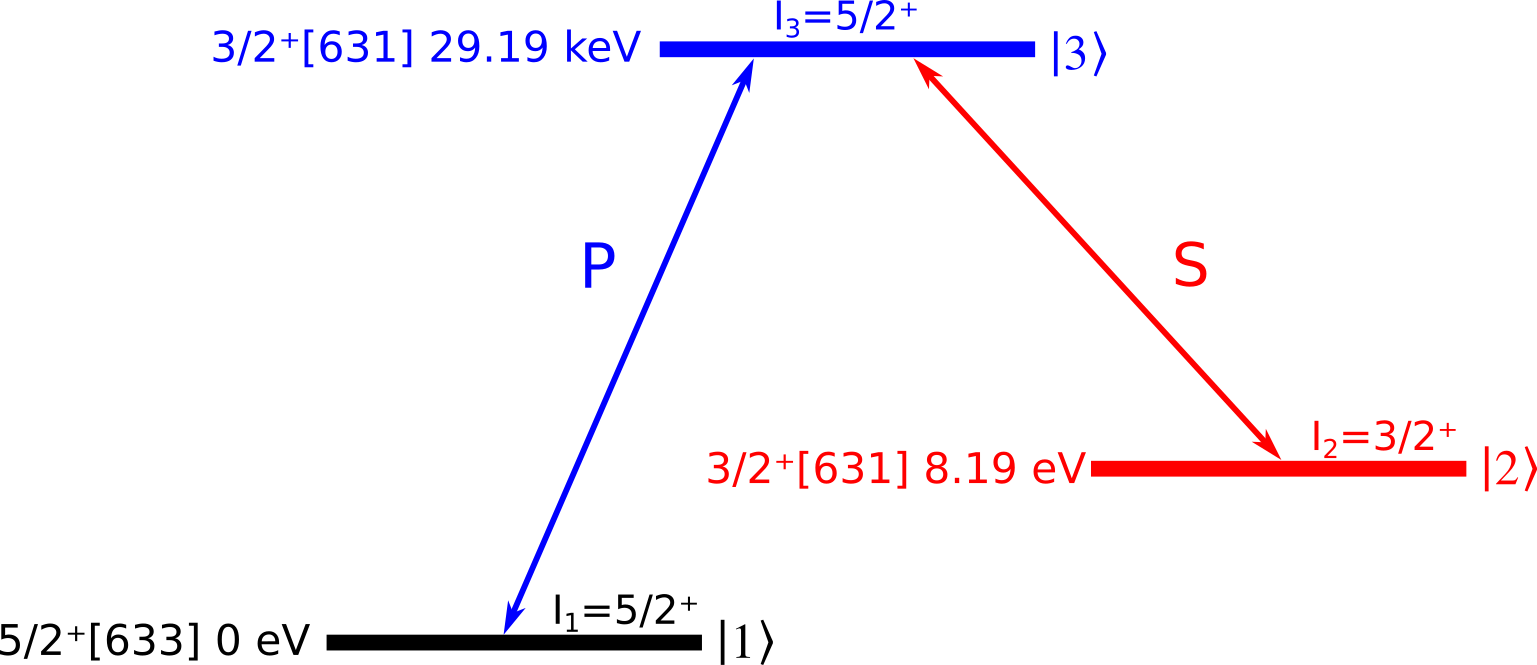}
    \caption{Three-level $\Lambda$ system comprising of the ground and  two first excited states of $^{229}$Th. Each state is labelled with its angular momentum, parity, Nilsson quantum numbers \cite{Nilsson1955} and energy. The letters P/S correspond to the pulse type (pump/Stokes) coupling to the respective levels. Energies are not to scale.}
    \label{3lvl}
\end{figure}

In this work we theoretically investigate  and optimize excitation schemes for the isomer via the 29.190 keV level taking advantage of coherence-based schemes known from  quantum optics and adapted for x-ray quantum optics \cite{Adams2013}. We consider the  $^{229}$Th three-level system of $\Lambda$-type (named so because of the level scheme reminiscent of the upper case Greek letter  $\Lambda$)
 comprising the ground ($|1\rangle$), isomeric ($|2\rangle$) and second excited ($|3\rangle$) states as illustrated in Fig.~\ref{3lvl}. We study efficient nuclear coherent population transfer (NCPT) from the ground state to the 8.19 eV isomeric state via two quantum optical transfer schemes: (i) 
Stimulated Raman adiabatic passage (STIRAP)\cite{bergmann_coherent_1998,bergmann_perspective_2015, Vitanov2017, Bergmann_2019}, and (ii) two subsequent $\pi$-pulses  \cite{scully_zubairy_1997} which pump the entire population first from $|1\rangle \rightarrow |3\rangle$ and subsequently from $|3\rangle \rightarrow |2\rangle$. 
Provided sufficient x-ray intensity, these transfer schemes allow to place the entire population from the ground state into the isomeric state in a coherent and controlled manner. Prior theoretical studies have discussed the possibility of nuclear STIRAP in the context of high-energy gamma-ray transitions \cite{Liao2011,Liao2013,iran2016,iran2017}.
 The peculiarities of the investigated $^{229}$Th  three-level system are that the two transitions to the upper state have low energy and narrow resonances, are almost degenerate in energy and have dominating internal conversion decay channels, with non-negligible multipole mixing. 

%%%%%%%%%%%%%%%%%%%%%%%%%%%%%%%%%%%%
%%%%%%%%%%%%%%%%%%%%%%%%%%%%%%%%%%%

We model the three-level-$\Lambda$-system within the density matrix approach and solve the master equation to determine the population transfer to the isomeric state. 
As the experimental setup involving two coherent x-ray pulses is not trivial, we
envisage three different scenarios for NCPT.  
The first scenario involves the interaction of  highly charged $^{229}$Th ions circulating at relativistic speed in a storage ring, with two UV laser fields driving the two x-ray transitions in the nuclear rest frame \cite{budker2021}. This scenario was proposed at the future Gamma Factory (GF) facility at CERN \cite{krasny2015gamma,budker2020,budker2021}.  Similarly, the second scenario discusses the interaction with coherent x-rays from an x-ray free electron laser (XFEL) \cite{xfel_overview,pellegrini2016,ALTARELLI20112845, nam2019,LCLSII} combined with moderately relativistic $^{229}$Th ions in a storage ring. The third  scenario investigates the interaction of a  $^{229}$Th solid-state target with coherent and resonant x-rays  ($E_{\lambda} \geq \SI{29.19}{\kilo \electronvolt}$) from a cavity-based x-ray source, for instance the x-ray free electron laser oscillator (XFELO) \cite{XFELO_Prop, xfeloAdams,qin, lindberg2011}. Our results identify the most feasible scenario from the three and investigate the specific requirements for each  NCPT mechanism, in particular the required laser intensities for efficient population of the isomer. 

The NCPT theoretical calculations require knowledge of nuclear transition parameters which are not experimentally available at the moment. In particular, the nuclear reduced transition probabilities $B(M1)$ for the magnetic dipole channel of both  $|3\rangle \rightarrow |1\rangle$ and  $|3\rangle \rightarrow |2\rangle$ transitions are an input for the calculation. We use our recently introduced nuclear structure model  \cite{Minkov_Palffy_PRL_2017,Minkov_Palffy_PRL_2019, Minkov_2021} to calculate these values and discuss the choice of our model parameters. Furthermore, we compare the newly obtained values for $B(M1)$ and related quantities with experimental data  available from two recent experiments \cite{masuda,Sikorsky2020}. While theory and experiment agree reasonably well on the $B(M1)$ values, our analysis reveals that the three experimentally available quantities are not completely consistent. We discuss the possible
values for the  radiative and total branching ratios which are of interest also for other experiments.

The paper is structured as follows. The theoretical model for the quantum optical transfer schemes to achieve NCPT is introduced in Sec.~\ref{theo}. 
Section \ref{param} introduces the nuclear transition properties of the three level-$\Lambda$ system and briefly discusses the nuclear model input used to obtain them. Furthermore, our analysis on the available experimental data is presented here. Our numerical results are given and discussed in Sec.~\ref{numres}. The paper concludes with a brief discussion in Sec.~\ref{concl}.

\section{Quantum optical transfer schemes \label{theo}}
\begin{figure}[!htp]
    \centering
    \includegraphics[width = 8
    cm]{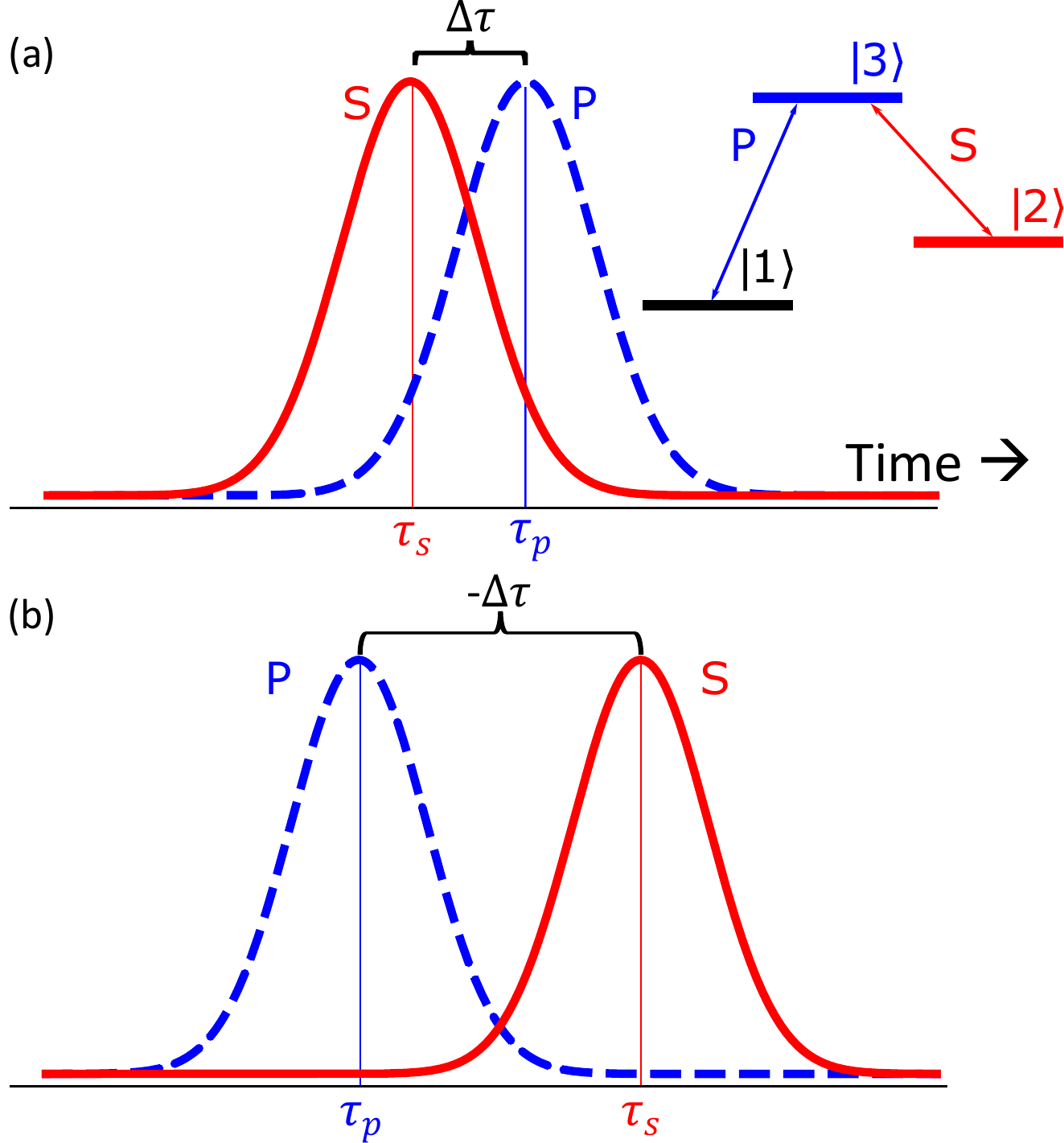}
    \caption{a) Generic pulse sequence for  STIRAP. $\Delta \tau = \tau_p-\tau_s$ indicates the delay between the pulses. b) Generic pulse sequence for NCPT via $\pi$-pulses.}
    \label{sequence}
\end{figure}

This Section introduces the theoretical models for the STIRAP and $\pi$-pulses quantum optical transfer schemes for our $\Lambda$ three-level system in $^{229}$Th illustrated in  Fig.~\ref{3lvl}. 
%\subsection{Quantum optical transfer schemes}
%%%%%%%%%%%%%%%%%%%%%%%%%%%%%%%%%%%%%%%%%%%%%%%%%%%%%
We investigate the interaction of two resonant fields driving our nuclear $\Lambda$ three-level system and the resulting change in population. NCPT can be achieved by means of STIRAP and two $\pi$-pulses. What distinguishes the two schemes is the arrival sequence of the two pulses. For STIRAP, the first arriving pulse is the one 
driving the transition $\ket{2} \leftrightarrow \ket{3}$, called the Stokes pulse. The pump pulse follows, partially overlapping with the Stokes pulse, and drives the second transition $\ket{1} \leftrightarrow \ket{3}$. In the two $\pi$-pulses scheme, the pump pulse arrives first, followed by the Stokes pulse. 
The two pulse sequences are illustrated for generic Gaussian-shaped pulses in Fig.~\ref{sequence}. 

%%%%%%%%%%%%%%%%%%%%%%%%%%%

%%%%%%%%%%%%%%%%%%%%%%%%%%%

STIRAP coherently transfers population between the two ground states of a $ \Lambda$-type system independently of the branching ratio or decay channels of the upper state  \cite{bergmann_coherent_1998, bergmann_perspective_2015}. 
The preceding Stokes pulse couples to the unoccupied states $\ket{2}$ and $\ket{3}$ and creates a coherent superposition thereof. Consequently, the overlapping pump pulse couples to the fully occupied ground state $\ket{1}$ and the pre-built coherence of $\ket{3}$ and $\ket{2}$. In this process a dark state \cite{bergmann_coherent_1998}
\begin{equation}
    \ket{\text{D}}=\frac{\Omega_s \left(t \right)}{\sqrt{\Omega_s \left(t \right)^2 +\Omega_p \left(t \right)^2}}\ket{1}- \frac{\Omega_p \left(t \right)}{\sqrt{\Omega_s \left(t \right)^2 +\Omega_p \left(t \right)^2}}\ket{2},
\end{equation}
is formed. Here, $\Omega_{s/p}$ are the time dependent Rabi frequencies \cite{palffy2007_PRC,Liao2011} of the Stokes and Pump pulse, respectively. By optimization of the laser parameters such as intensity or delay time $\Delta \tau$ between the Stokes and pump pulses, one is able to transfer the entire population from $\ket{1}$ to $\ket{2}$ via the dark state, without 
occupying the intermediate state $\ket{3}$.
A rather empirical condition for each pulse, namely the adiabatic condition \cite{VITANOV2001,kuhn}
\begin{equation}
\label{adiabaticcond}
     \Omega_0^2 \geq \frac{2\ln{2} \cdot 100 }{t_{\text{pul}}^2}
\end{equation}
provides a guideline for the optimization.
In this equation $\Omega_0$ denotes the Rabi frequency amplitude of the respective pulse and $t_{\text{pul}}$ the pulse duration at FHWM, respectively.
Note that this formula only holds if the pulses are fully temporally coherent and have no phase fluctuations.
For simplicity, we will assume throughout this paper that no (relative) phase fluctuations occur.

For the $\pi$-pulse technique, the pulse sequence is opposite to STIRAP. Thereby, the population is initially transferred from $\ket{1}$ to $\ket{3}$ by the pump pulse and from there immediately to $\ket{2}$ by the Stokes pulse. To transfer the entire population into the desired state each pulse should satisfy \cite{scully_zubairy_1997}
 \begin{equation}
 \label{pi}
     \int_{-\infty}^{\infty} \Omega \left( t\right) \text{d}t = \pi\, .
 \end{equation}
By solving the integral, we arrive to an expression for the pulse requirements to achieve transfer rates of unity.
For Gaussian pulses we obtain the expression 
\begin{equation} \label{pi-cond}
    \Omega_0^2 =2 \ln{2} \frac{\pi}{t_{\text{pul}}^2}\, .
\end{equation}

Both methods are a challenge to the experiment, since the transfer criteria in Eqs.~\eqref{adiabaticcond} and~\eqref{pi} require very large laser intensities for the available pulse durations. 
However, implementing such coherence-based schemes has the advantage that the population transfer is no longer dependent on incoherent processes like spontaneous decay.
If the lifetime of the intermediate state $\ket{3}$ is shorter than the radiative coupling, then STIRAP is the method of choice. Apart from that, the $\pi$-pulse method is applicable. 
An advantage of the $\pi$-pulse method is that one has a larger window for the pulse delay time, which is only constrained by the lifetime of the intermediate state $\ket{3}$. Furthermore, the $\pi$-pulse method requires smaller intensities than STIRAP. However, STIRAP is more robust than the $\pi$-pulses method as far as 
parameter variations are concerned \cite{bergmann_coherent_1998}, as long as the temporal coherence of the pulses is secured.

\subsection{Density matrix approach}
%%%%%%%%%%%%%%%%%%%%%%%%%%%%%%%%%%%%%%%%%%%
We model the population of the three-level-$\Lambda$ system within the density matrix approach. The density matrix is defined as
\begin{equation}
  \rho \left( t\right) = \sum_{i,j} \rho_{ij}\left( t\right) \ket{i}\bra{j}  
\end{equation}
where $\{i,j\}\in\{1,2,3\}$. The diagonal elements $\rho_{ii}$ denote the level-population and the off-diagonal elements $\rho_{ij}$ the coherences, respectively.
Our starting point to compute the population transfer in $^{229}$Th is the master equation \cite{scully_zubairy_1997, palffy2007_PRC}
\begin{equation}
    \partial_t \rho \left( t\right) = \frac{1}{i \hbar} \left[ \mathcal{H}\left( t\right), \rho \left( t\right) \right] +\rho_{\text{relax}}\left( t\right),
\end{equation}
where $\mathcal{H}\left( t\right)$ denotes the interaction Hamiltonian and $\rho_{\text{relax}}\left( t\right)$ the relaxation term, respectively. Furthermore, $\hbar$ denotes the reduced Planck constant.  
The Hamiltonian reads \cite{Liao2011, Liao2013}
\begin{equation}
    \mathcal{H}\left( t\right) = -\frac{\hbar}{2}
    \begin{pmatrix}
      0 & 0& \Omega_p^\ast\left( t\right)\\
     0& -2 \left( \Delta_p - \Delta_s \right) &  \Omega_s^\ast\left( t\right)\\
     \Omega_p\left( t\right)& \Omega_s \left( t\right) & 2 \Delta_p
    \end{pmatrix}.
\end{equation}
Here,  $\Omega_{p/s}$ is the time dependent Rabi-frequency  and $\Delta_{p/s}$ the detuning of the pump$/$Stokes field, respectively. The expressions for the Rabi frequencies are given below. 
The relaxation matrix embodying several decay channels  has the form
\begin{equation}
    \rho_{\text{relax}}\left( t\right) = \frac{1}{\hbar}\begin{pmatrix}
    \text{BR}_{31} \Gamma \rho_{33}  & 0 & -0.5\Gamma\rho_{13} \\
      0&  \text{BR}_{32} \Gamma \rho_{33} & -0.5\Gamma \rho_{23} \\
     -0.5\Gamma \rho_{31} & -0.5\Gamma\rho_{32}& -\rho_{33} \Gamma \label{eq:rho_relax}
    \end{pmatrix}.
\end{equation}
where $\text{BR}_{3j}$ denotes the branching ratio for the transition $\ket{3}\rightarrow \ket{j}$ where $j\in \{1,2\}$ and $\Gamma$ is the total decay rate of $\ket{3}$, respectively.
For simplicity we assume throughout this work that
both laser fields are fully temporally coherent during the radiative coupling. In addition, we neglect the decay of the isomeric level, since the lifetime of $\ket{2}$ is much larger than the lifetime of $\ket{3}$.
If not mentioned otherwise, initially the nuclear population is in the ground state such that $\rho_{11}\left(0\right)=1$ and $\rho_{22}\left(0\right)=\rho_{33}\left(0\right)=0$.

%..............     CONTINUE HERE!................................................

\subsubsection{Lab frame}
%%%%%%%%%%%%%%%%%%%%%%%%%%%%%%%%%%%%
The time dependent Rabi frequencies in the laboratory frame are assumed to be Gaussian, such that 
\begin{equation}
    \Omega_{p \left( s\right)} = \Omega_{0,p \left( s\right)}\exp\left(-2\ln{2} \frac{\left(t-\tau_{p \left( s\right)}\right)^2}{t_{\text{pul}^2}} \right)
\end{equation}
where $\Omega_{0,p \left( s\right)}$ denotes the Rabi amplitude and  $\tau_{p \left( s\right)}$ the temporal peak position of the pump$/$Stokes field, respectively.
We can express the Rabi amplitude for the nuclear transition $\ket{i} \rightarrow \ket{j}$ as \cite{Liao2011,Liao2013}
\begin{equation}
\begin{split}
        \Omega_{0,ij} = \sqrt{\frac{16 \pi I_0}{c \epsilon_0 \hbar^2}}\sqrt{\frac{\left(2I_{i}+1 \right) \left(L_{ij} + 1 \right)  }{L_{ij}}} \\ \times \frac{k_{ij}^{L_{ij}-1}}{\left(2L_{ij} +1 \right)!! } \sqrt{B_{ij} \left(\mu L_{ij}\right)} = \sqrt{I_0} \xi_{ij} \label{eq:Rabi-freq}
\end{split}
\end{equation}
where $I_0$ indicates the peak intensity of the radiation field, $c$ the speed of light, $\epsilon_0$ the vacuum permittivity, $I_i$ the nuclear spin of level $\ket{i}$, $L_{ij}$ the multipolarity, $k_{ij}$ the wave number and $B_{ij} \left(\mu L_{ij}\right)$ the reduced transition probability of the  transition, respectively.
The index $\mu$ corresponds to the radiation multipole type electric or magnetic $\mu \in \{E, M \}$.
For the purpose of our calculation of radiative couplings, we only consider the first dominant multipole order in our Rabi frequency, which is the $M1$ channel.

With this concrete expression for the Rabi frequency, we can rewrite the transfer criteria in Eq.~\eqref{adiabaticcond} and Eq.~\eqref{pi} in terms of radiation intensity. We then obtain 
\begin{equation}
I_0 \geq \frac{2 \ln{2} \cdot 100}{t_{\text{pul}}^2 \xi_{ij}^2} \label{int-stirap}
\end{equation}
for STIRAP and
\begin{equation}
    I_0 = \frac{2\ln{2} \cdot\pi}{t_{\text{pul}}^2 \xi_{ij}^2} \label{int-pi}
\end{equation}
for the $\pi$-pulses, respectively.
We use these expressions to compute the ideal intensity required to transfer the entire population from the ground state to the desired isomeric state with hardly any losses.

\subsubsection{Ion rest frame }
%%%%%%%%%%%%%%%%%%%%%%%%%%%%%%%%%%%%%%%%%%%%%
In case of the relativistic acceleration of the nuclei, the laser photon parameters are relativistically boosted in the rest frame of the ion according to 
\begin{equation}
    \omega_3 - \omega_{1(2)} = \gamma\left (1 + \beta \cos{\theta} \right) \omega_{P(S)}\, ,
\end{equation}
where $\gamma$ is the relativistic Lorentz factor, $\beta = v/c$, with $v$ the ion velocity, and $\theta$ denotes the impact angle between radiation and nuclei, respectively.
Thus, the pulse width and the peak intensity change in the nuclear rest frame as \cite{Buervenich2006}
\begin{align}
    t_{\text{pul}} \rightarrow \frac{t_{\text{pul}}}{\gamma \left(1+ \beta  \right)}\, , \\
    I_0 \rightarrow  I_0 \gamma^2\left(1+ \beta\right)^2.
\end{align}
Here, we have assumed that the interaction between the ion bunch and the laser pulses is collinear ($\theta \approx 0$).
Furthermore, as the ions approach the speed of light, the relativistic factor $\gamma \left(1+ \beta \right)\rightarrow 2 \gamma$.
With the above considerations the time dependent Rabi frequency in the nuclear rest frame becomes
\begin{equation}
    \Omega_{p \left( s\right)}^{\text{Rest}} = 2\gamma  \Omega_{0,p \left( s\right)}^{\text{Lab}}\exp\left(-2\ln{2} \frac{4 \gamma^2\left(t-\tau_{p \left( s\right)}\right)^2}{t_{\text{pul}}^2} \right)\, .
\end{equation}
Also in the ion rest frame we can rewrite the adiabaticity criterion for the boosted intensities in analogy with Eqs.~\eqref{int-stirap} and \eqref{int-pi}.

\section{The $^{229}$Th Three-Level-$\Lambda$ system \label{param}}
In this Section we discuss our present knowledge on the nuclear transition properties of the  $^{229}$Th three-level-$\Lambda$ system under investigation. 
From the nuclear structure point of view, the 29.19 keV state with
angular momentum $I=5/2^{+}$ and the isomeric state with
$I=3/2^{+}$ belong to the excited (non-yrast) band built on the
$K^\pi [Nn_z\Lambda]=3/2^+ [631]$ single-neutron orbital, with $K$  the
projection of the angular momentum on the intrinsic nuclear
symmetry axis, $\pi$ the parity and $N$,  $n_z$ and $\Lambda$ the Nilsson
asymptotic quantum numbers \cite{NR95}, respectively.
%From the nuclear structure point of view, the 29.19 keV $5/2^+\, [631]$ and the isomeric $3/2^+\, [631]$ states belong to the excited (non-yrast) band.
Thus, 
the transition $\ket{1} \leftrightarrow \ket{3}$ corresponds to the cross-band (cr) transition, while the transition $\ket{2} \leftrightarrow \ket{3}$ corresponds to the in-band (in) transition.
For laser excitation, the radiative channels of the two transitions are of interest, with the reduced transition probabilities $B(M1)$ entering the Rabi frequency amplitudes in Eqs.~\eqref{eq:Rabi-freq}. However, the relaxation matrix \eqref{eq:rho_relax} contains also other quantities, such as incoherent decay rates, which could include also the internal conversion (IC) channel, and several types of branching ratios. Thus, we are interested in $B(M1)$ and $B(E2)$ values for both in-band and cross-band transitions, as well as the related IC coefficients,  and the resulting branching ratios $\mathrm{BR}_{31}$ and $\mathrm{BR}_{32}$.  
The corresponding transition rates are only partially known from experiments. Alternatively, we can obtain these quantities from a nuclear structure model which was so far successfully used to predict the low-lying level structure of $^{229}$Th \cite{Minkov_Palffy_PRL_2017,Minkov_Palffy_PRL_2019,Minkov_2021}.
In the following we present our nuclear structure results on the transition properties of the in-band and cross-band transition. We discuss these values in comparison with available experimental data and check the consistency of the available parameter sets.

\subsection{Theoretical predictions}
%%%%%%%%%%%%%%%%%%%%%%%%%%%%%%%%%%%%%%%%%%

The actinide nuclei and in particular the even-odd isotopes among them present a rich nuclear structure. A model approach capable to incorporate the shape-dynamic properties together with the intrinsic structure characteristics typical for the actinide nuclei has been under development in the last decade~\cite{b2b3mod,b2b3odd,qocsmod,WM10,MDSS09,MDSS10, NM13}.  It considers a collective quadrupole-octupole (QO) vibration-rotation motion of the nucleus which in the particular case of odd-mass nuclei is coupled to the motion of the single (odd) nucleon within a reflection-asymmetric deformed potential. The collective motion is described through the so-called coherent QO mode (CQOM) giving raise to the quasi parity-doublet structure of the spectrum~\cite{b2b3mod,b2b3odd}, whereas the single-particle (s.p.) one is determined by the deformed shell model (DSM) with reflection-asymmetric Woods-Saxon potential~\cite{qocsmod} and pairing correlations of Bardeen-Cooper-Schrieffer (BCS) type included as in Ref.~\cite{WM10}. The Coriolis interaction between CQOM and the odd nucleon was originally considered in~\cite{MDSS09,MDSS10}, whereas the effect of Coriolis decoupling and $K$-mixing on the rotation-vibration was taken into account in Ref.~\cite{NM13}. 
% levels, with $K$ being the projection of the angular momentum on the intrinsic nuclear symmetry axis, was taken into account in~\cite{NM13}. 

All the model aspects outlined above have been assembled together in Ref.~\cite{Minkov_Palffy_PRL_2017} in a detailed nuclear-structure-model description of the low-lying positive- and negative-parity excited levels and transition probabilities observed in $^{229}$Th. This allowed the description of several experimentally known $B(M1)$ and $B(E2)$ reduced probabilities and prediction of the two unknown isomeric ones, responsible for the radiative decay of the 8.19 eV $K^\pi=3/2^{+}$-isomer to the $K^\pi=5/2^{+}$ ground state,  corresponding to the cross-band transition $\ket{1} \leftrightarrow \ket{2}$ in the scheme of Fig.~\ref{3lvl}.

The two states are considered as almost degenerate quasi-particle bandheads with a superposed collective QO vibration-rotation mode giving raise to yrast $K^\pi=5/2^{+}$ and non-yrast $K^\pi=3/2^{+}$ quasi parity-doublet structures. The isomer decay is obtained as the result of a Coriolis mixing emerging from a remarkably fine interplay between the coherent QO motion of the core and the single-nucleon motion within  the  reflection-asymmetric deformed potential. Despite earlier statements on the weakness of the Coriolis mixing~\cite{Dyk98,Tkalya15}, we found that only because  of the Coriolis  $K$-mixing interaction  can we  explain the presence of the  $K$-forbidden $M1$ and $E2$ transitions between the yrast and non-yrast bands, otherwise  forbidden due to the overall axial symmetry of the problem. The same holds also for the cross-band $|3\rangle \rightarrow |1\rangle$ transition of the three-level system investigated here, which can only be accounted for by the inclusion of Coriolis mixing in the model.

Within this model it is also clear that the two electromagnetic multipole contributions have different  origins. The $E2$ transition is mainly related to the collective part, whereas the $M1$ component emerges from the  single-nucleon degree of freedom~\cite{Ring1980}.  Nevertheless, the collective QO mode has a strong indirect influence on the $M1$ transition via the s.p. coupling to the nuclear core. Vice-versa, the collective part is decisive for the $E2$ transition, however with indirect influence from the single nucleon via the particle-core coupling. The very fine balance between the different degrees of freedom and its role in the formation of the total dynamics of the nucleus and its isomer properties including energy, electromagnetic transition rates and magnetic moments were examined in detail in \cite{Minkov_2021}. It was confirmed that the reasoning for the existence and the decay properties of the $^{229m}$Th state is strongly related to all nuclear structure model ingredients originally considered in Ref.~\cite{Minkov_Palffy_PRL_2017}, namely, the collective core, the single-nucleon motion in the deformed potential and the Coriolis interaction. 

 A specificity of the model in calculating the $M1$ reduced transition probability is the use of an additional input related to the intrinsic-spin and collective gyromagnetic factors, which are attenuated by multiplication with the respective quenching factors $q_s$ and $q_R$, as explained in Refs. \cite{Minkov_Palffy_PRL_2019,Minkov_2021}. The nuclear-structure grounds and theoretical motivation for the introduction of the two quenching factors are discussed in Ref. \cite{Minkov_Palffy_PRL_2019} based on the original adjustment of the model parameters to the $^{229}$Th energy levels and transition rates reported in \cite{Minkov_Palffy_PRL_2017}. The dependence of the predicted $^{229m}$Th electromagnetic properties on the $q_s$ and $q_R$ quenching factors and QO deformations is examined in Ref. \cite{Minkov_2021} through different fits including in addition the  observed ground- and isomeric- state magnetic dipole moments. Physically consistent model descriptions of energy, transition rates and magnetic moments are obtained in a narrow QO-deformation region by varying $q_s$ between 0.6 and 0.55 and $q_R$ between 0.6 and 0.45, with corresponding minimal (and smooth) changes in the adjusted CQOM and Coriolis-mixing parameters, and with the BCS parameters being fixed as in Refs.~\cite{Minkov_Palffy_PRL_2017,Minkov_Palffy_PRL_2019} (see Figs.~6-12 in \cite{Minkov_2021}).

So far the in-band and cross-band decay transitions of the 29.19 keV level were not considered in our works  \cite{Minkov_Palffy_PRL_2017,Minkov_Palffy_PRL_2019,Minkov_2021}. Here we present our (new) theoretical predictions for the $B(M1)$ and $B(E2)$ values in this level obtained through the quenching factors and other model parameters as explained above.
Table  \ref{Th229ParamPRL19} presents the $B(M1)$ values obtained through the CQOM-DSM-BCS and Coriolis-mixing parameters used in Ref.~\cite{Minkov_Palffy_PRL_2019} for $q_s=0.6$ and $q_s=0.55$ with several values of the collective gyromagnetic quenching factor $q_R$ between 1 and 0.45. The corresponding values of the $B(E2)$ reduced transition probabilities, which are independent of the two quenching factors, are $B(E2;|3\rangle \rightarrow |1\rangle)=27.11$ W.u. (Weisskopf units) and  $B(E2;|3\rangle \rightarrow  |2\rangle )=239.18$ W.u.

Table \ref{Th229ParamPRC21} presents the $B(M1)$ and $B(E2)$ values obtained at few pairs of $q_s$ and $q_R$ values through the corresponding sets of CQOM and Coriolis-mixing parameters adjusted in Ref.~\cite{Minkov_2021} taking into account the experimental magnetic moments with the DSM and BCS parameters kept the same as in Ref.~\cite{Minkov_Palffy_PRL_2019}. We note that here the reduced transition probabilities for the $E2$ channel slightly vary with $q_s$ and $q_R$ due to the (slightly) different CQOM and Coriolis-mixing parameter values obtained in the different fits.
%%
%For comparison, the results in Table \ref{Th229ParamPRC21} are calculated using several model parameter sets from Ref.~\cite{Minkov_2021}. The reduced transition probabilities for the $E2$ channel are vary only slightly (due to the different CQOM model parameters) and are also presented
 A comparison between Table \ref{Th229ParamPRL19} and Table \ref{Th229ParamPRC21} shows that for the same set of quenching factors, the reduced transition probabilities present some variations depending on the chosen CQOM and Coriolis mixing model parameters. Overall, the in-band transition $ |3\rangle \rightarrow |2\rangle $ presents much stronger reduced transition probabilities for both multipolarities than the cross-band transition $ |3\rangle \rightarrow |1\rangle $.

%%%%%%%%%%%%%%%%%%%%%%%%%%%%%%%%%%%%%%%%%%%%%%%%%%%%%%%%%
\renewcommand{\arraystretch}{1.3} 
\begin{table}[]
    \centering
    \begin{tabular}{llcc}
    \hline
    \hline
    
    \multirow{2}{4em}{  $q_S$ }  & \multirow{2}{4em}{  $q_R$ }   & \multicolumn{2}{c}{$B(M1)$ (W.u.)} \\ \cline{3-4}
     &  &  $ |3\rangle \rightarrow  |1\rangle$   & $ |3\rangle \rightarrow |2\rangle $   \\
    \hline
    0.6    & 1 & 0.0012 &0.0648 \\
       & 0.8 & 0.0020  &0.0544 \\
        & 0.7  & 0.0025  & 0.0495\\
        & 0.6 & 0.0030  & 0.0449 \\
        &  0.5& 0.0036  &0.0405\\
        & 0.45 & 0.0039  &0.0383\\
        \hline
       0.55 & 0.5 & 0.0028  & 0.0357 \\
        & 0.45& 0.0031  &0.0337\\
        \hline 
        \hline
    \end{tabular}
    \caption{Predicted $B(M1)$ values  obtained for several spin and gyromagnetic quenching factors $q_s$ and $q_R$ with all other model  parameters taken from Ref.~\cite{Minkov_Palffy_PRL_2019}. See text for further explanations.}
    \label{Th229ParamPRL19}
\end{table}

\renewcommand{\arraystretch}{1.2} 
\begin{table}[]
    \centering
    \begin{tabular}{llccccc}
    \hline
    \hline
    \multirow{2}{4em}{  $q_S$ }  & \multirow{2}{4em}{  $q_R$ }   & \multicolumn{2}{c}{$B(M1)$ (W.u.)} & & \multicolumn{2}{c}{$B(E2)$ (W.u.)} \\ \cline{3-4} \cline{6-7}
     &  &  $ |3\rangle \rightarrow  |1\rangle$   & $ |3\rangle \rightarrow |2\rangle $  & &  $|3\rangle \rightarrow  |1\rangle$   & $ |3\rangle \rightarrow |2\rangle $ \\
    \hline
    0.6 & 0.6 & 0.0043  & 0.0432 & & 39.49 & 234.86 \\
   & 0.5 & 0.0050 & 0.0390 & & 38.23 & 235.52 \\
   \hline 
    0.55   & 0.5 & 0.0035 & 0.0348 &&  34.19 & 235.87\\
       & 0.45 & 0.0035 & 0.0332 & & 31.44 & 236.11\\
      
        \hline 
        \hline
    \end{tabular}
    \caption{Predicted $B(M1)$ and $B(E2)$ values  obtained for several spin and gyromagnetic quenching factors $q_s$ and $q_R$ with all other model  parameters taken from Ref.~\cite{Minkov_Palffy_PRL_2019}. See text for further  explanations.}
    \label{Th229ParamPRC21}
\end{table}

\subsubsection{Radiative rates}
%%%%%%%%%%%%%%%%%%%%%%%%%%%%%%%%%%%%%
The reduced transition probabilities give access to the radiative transition rates which can be calculated according to \cite{Ring1980}
\begin{equation}
    T\left( M1\right) = \num{1.779e4}\cdot E[\si{\kilo \electronvolt}]^3 \cdot B \left(M1\right) \label{T-M1}
\end{equation}
 for $M1$ transitions and
 \begin{equation}
        T\left( E2\right) = \num{1.223e-6}\cdot E[\si{\kilo \electronvolt}]^5 \cdot B \left(E2\right)
        \label{radrateE2}
 \end{equation}
for $E2$ transitions, where the rates $T(\mu L)$ are given in s$^{-1}$, $E[\si{\kilo \electronvolt}]$ is the transition energy in keV, $B(M1)$ in $\mu_{N}^{2}$ with $\mu_N$ the nuclear magneton and $B(E2)$ in e$^{2}$fm$^{2}$, respectively. The total decay rate is obtained as a sum over the rates of the two multipolarities $M1$ and $E2$, and the corresponding width for the in-(cross-)band transitions is given by 
\begin{equation}
\label{brgamma}
    \Gamma_\gamma = \hbar \sum_{\mu L} T(\mu L) \, .
\end{equation}
We note that due to the suppressing factors in the expression \eqref{radrateE2}, the $E2$ contribution for the radiative decay is approx. two orders of magnitude smaller than the $M1$ one for both in- and cross-band transitions. Thus, the $E2$ channel can be neglected in Eq.~\eqref{brgamma}. However, this will not be the case for the corresponding IC rates, as it will be discussed in the following.

\subsubsection{IC rates \label{sec_ic}}
%%%%%%%%%%%%%%%%%%%%%%%%%%%%%%%%%%%%%%%%%%%%%%%
In the process of IC, the energy of the nuclear excited state is transferred to an atomic electron, which is kicked out of its shell. The nuclear excitation energy should therefore exceed the binding energy of the IC electron. IC rates can be calculated taking into account the interplay between atomic and nuclear degrees of freedom. However, since the nuclear part can be related to the reduced transition probabilities $B(\mu L)$, the quantity of interest is the IC coefficient $\alpha$, defined as the ratio of the IC and radiative rates. The IC coefficient contains the remaining atomic structure information and can be accurately calculated  with existing codes such as BrIcc \cite{Kibedi2008IC_coeff}. 

The IC coefficients are calculated according to the nuclear transition energy and the transition multipolarity. For the two considered transitions  $|3\rangle \rightarrow |2\rangle $ and $|3\rangle \rightarrow |1\rangle $, the transition energies differ by just 8 eV and the calculated IC coefficients are for all practical purposes identical for each considered multipolarity. The $E2$ IC coefficients are approx.~a factor 30 higher than the $M1$ coefficients, compensating for the smaller radiative rates of the former. Thus, the $E2$ IC channels are not negligible. Both  $E2$ and $M1$ IC coefficients are much larger than one, rendering IC the strongest decay channel for the 29 keV excited state. The total IC rate for the in-(cross-)band transition is given by the sum of the $M1$ and $E2$ channels, with the corresponding IC width 
\begin{equation}
\label{IC-width}
    \Gamma_{\mathrm IC} = \hbar \sum_{\mu L} \alpha(\mu L) T(\mu L) \, .
\end{equation}
In the following we use $\alpha\left(M1 \right)=151.07$ and  $\alpha\left(E2 \right)=4401.61$ for both in-band and cross-band transition. These values are interpolated from the IC coefficients tabulated in Ref.~\cite{ictable}.

\subsection{Analysis of available experimental data}
%%%%%%%%%%%%%%%%%%%%%%%%%%%%%%%%%%%%%
The experimental data on the transitions connecting the 29 keV level to the ground and isomeric states is scarce. There are only three related quantities that have been reported experimentally so far from two different experiments: (i) the cross-band radiative transition width reported in Ref.~\cite{masuda}, $\Gamma_\gamma^{\mathrm cr}=1.70\pm 0.40\, \si{\nano \electronvolt}$, (ii) from the same experiment, the total half-life of the state $|3\rangle$, dominated by the two IC decays to the isomer and ground states, $T_{1/2}=82.2\pm 4.0\, \si{\pico\second}$ and (iii) the radiative branching ratio of the cross-band transition, $\mathrm{BR}_{31}^\gamma=9.3(6)~\%$ \cite{Sikorsky2020}. 

From (i), we can deduce the corresponding $B(M1)$ for the cross-band transition. Neglecting the $E2$ multipole mixing (which, as discussed above, is a very good approximation), and using expression \eqref{T-M1} we obtain $B(M1;|3\rangle \rightarrow |1\rangle)=0.00326\pm 0.00076$ W.u. This value is in good agreement with most of the theoretical predictions listed in Tables \ref{Th229ParamPRL19} and \ref{Th229ParamPRC21}. Combining this value with the  measured branching ratio $\mathrm{BR}_{31}^\gamma$ reported in Ref.~\cite{Sikorsky2020}, we can extract the in-band reduced transition probability. Indeed, neglecting the multipole mixing in the radiative decay, we have
\begin{equation}
    \mathrm{BR}_{31}^\gamma = \frac{\Gamma_\gamma^{31}}{\Gamma_\gamma^{32}+\Gamma_\gamma^{31}}\, ,
\end{equation}
leading to $B(M1;|3\rangle \rightarrow |2\rangle)=0.0318$ W.u., within the uncertainty interval $[0.0227,0.0420]$ W.u. This value is slightly smaller than the theoretical predictions in Tables \ref{Th229ParamPRL19} and \ref{Th229ParamPRC21}. 

The total half-life of the upper state $|3\rangle$ depends on both radiative and IC decay channels of the two transitions, and here the multipole mixing needs to be taken into account. As a result, we have 
\begin{eqnarray}
    T_{1/2}&=&\ln{2}\large\{[1+\alpha(M1)][T_{31}(M1)+T_{32}(M1)] \nonumber \\
     &&+[1+\alpha(E2)][T_{31}(E2)+T_{32}(E2)]\}^{-1}\, . \label{thalf}
\end{eqnarray}
Despite reliable theoretical values for the two IC coefficients (the numerical values were given in Sec.~\ref{sec_ic}),  
this expression is not sufficient to determine the two remaining unknowns $T_{32}(E2)$ and $T_{31}(E2)$ and the corresponding reduced transition probabilities $B(E2)$. We note that mixing ratios for the two transitions are not available from experiments. Thus,  some further assumption starting from our theory knowledge is required.

Inspection of the theoretically predicted $B(E2)$ values shows that regardless of the used model parameters, the in-band reduced transition probability  $B(E2;|3\rangle \rightarrow |2\rangle )$ value is rather stable, with values varying between 234.86 W.u. and 239.18 W.u. In the following we will use the average value $B(E2;|3\rangle \rightarrow |2\rangle)=237.02$ W.u. to obtain $T_{32}(E2)$ and derive the remaining unknown rate $T_{31}(E2)$ from Eq.~\eqref{thalf}. For the $M1$ channels we use the values deduced above from the experimental observations, dropping the error bars. We obtain $T_{31}(E2)=4.478\cdot 10^5\si{\second}^{-1}$ and correspondingly 
$B(E2;|3\rangle \rightarrow |1\rangle)=207.65$ W.u. We note that this value is unexpectedly large for the cross-band transition, being of the same order of magnitude with the in-band reduced transition  probability. In case we start from a theoretical value for $B(E2;|3\rangle \rightarrow |1\rangle)$ in the range shown in Table \ref{Th229ParamPRC21}, we obtain for the in-band transition an unexpectedly large reduced transition probability of $B(E2;|3\rangle \rightarrow |2\rangle)>400$ W.u.
These inconsistencies suggests that the experimental values obtained from the two experiments do not match exactly. We have checked that a consistent set of results can be obtained
if in expression~\eqref{thalf} we use values close to or at the
upper uncertainty limits of the experimental values of the total
half-life of state $|3\rangle$ and the two $B(M1)$ transition rates,
such as $T_{1/2}=85.6$~ps, $B(M1;|3\rangle \rightarrow
|1\rangle)=0.0040$ W.u. and $B(M1;|3\rangle \rightarrow
|2\rangle)=0.0420$ W.u. This may be considered as an indication for
the need of further more precise experimental determination of
these quantities.

We note that also Ref.~\cite{masuda} arrives at inconsistencies when combining the reported experimental lifetime and cross-band transition width with theoretical values from Refs.~\cite{KrogerReich1976,Barci2003}. The latter theoretical values are obtained from the standard rotational model and are not expected to be particularly accurate for $^{229}$Th, especially in view of the most recent insights into the nuclear structure origins of the Th isomer \cite{Minkov_Palffy_PRL_2017,Minkov_Palffy_PRL_2019,Minkov_2021}. By combining the experimental $T_{1/2}$ and $\Gamma_\gamma^{\mathrm cr}$ with theoretical predictions on the in-band radiative decay \cite{Barci2003}, the authors of Ref.~\cite{masuda} deduce a total cross-band IC coefficient of 1,370. This value hints to an implausible multipole mixing for the cross-band transition. Even by using the (equally implausible) value $B(E2);|3\rangle \rightarrow |1\rangle)=207.65$ W.u. inferred above, the cross-band IC coefficient would be a factor of two smaller. A closer inspection of the theoretical values in Ref.~\cite{Barci2003} shows that the predicted cross-band transition width disagrees with the experimental value by a factor 2. The rotational model predictions are therefore not suitable for combining experimental and theoretical results to deduce correct branching ratios and related transition parameters. This inconsistency reflects in the unexpectedly large IC coefficient value.

\subsection{Choice of nuclear transition parameters} 
%%%%%%%%%%%%%%%%%%%%%%%%%%%%%%%%%%%%%%%
For our STIRAP calculations in Th ions or atoms we require the $M1$ reduced transition probabilities for the in- and cross-band transitions, the total decay rate of level $|3\rangle$ and the branching ratios $\mathrm{BR}_{31}$ and $\mathrm{BR}_{32}$ with and without including the IC channel. In the following we list in Table ~\ref{Used-param} the used sets of nuclear transition parameters and comment on the reliability of our choice. 

%%%%%%%%%%%%%%%%%%%%%%%%%%%%%%%%%%%%%%
\begin{table}[htpb]
    \centering
    \begin{tabular*}{\linewidth}{@{\extracolsep{\fill}} llccc}
    \hline 
    \hline
  &  & cross-band & in-band & total decay \\
  & & $|3\rangle \rightarrow |1\rangle$  & $|3\rangle \rightarrow |2\rangle$ & of $|3\rangle$ \\
    \hline 
\multirow{2}{4em}{$B(M1)$ [W.u.]} & theo\footnote{Values are obtained using quenching factors $q_s=q_R=0.6$ and the model parameters from Ref.~\cite{Minkov_Palffy_PRL_2019}.} & 0.0030 & 0.0449& \\
    
    & exp\footnote{Values obtained combining the measured cross-band radiative width \cite{masuda} and radiative branching ratio \cite{Sikorsky2020} values.} & 0.00326(76) & $0.0318\substack{+ 0.0102 \\ -0.0091}$ & \\     
    & & & & \\
    \multirow{2}{4em}{BR$_\gamma$[\%]} & theo\footnote{Obtained with the theoretical $B(M1)$ values listed in this Table.} &  6.3 & 93.7  &  \\   
                                            & exp\footnote{From Ref.~\cite{Sikorsky2020}.} & 9.3(6) &90.7(6) & \\ 
                           
& & & & \\                                            
     \multirow{2}{4em}{BR[\%]} & theo &  7.5 & 92.5  &  \\   
                                            & exp\footnote{Using the reduced transition probabilities extracted from the experimental data discussed above, i.e., $B(M1;|3\rangle \rightarrow |1\rangle)=0.00326$ W.u., $B(M1;|3\rangle \rightarrow |2\rangle)=0.0318$ W.u., $B(E2);|3\rangle \rightarrow |1\rangle)=207.65$ W.u. and the theoretical assumption $B(E2);|3\rangle \rightarrow |2\rangle)=237.02$ W.u.} & 28.01 & 71.99 & \\ 
& & & & \\                                            
                                            
   \multirow{2}{4em}{$\Gamma_\gamma$[neV]} & theo$^c$ & 1.6 & 23.71 & 25.31 \\   
                                            & exp\footnote{the cross-band value was reported in Ref.~\cite{masuda}; the other two values are obtained using in addition the experimental branching ratio reported in Ref.~\cite{Sikorsky2020}.}  & 1.7(4) & $16.56\substack{+ 5.328 \\ -4.734}$ & $18.26\substack{+ 5.728 \\ -5.134}$ \\
                                            
& & & & \\                                                   
       \multirow{2}{4em}{$\Gamma$[$\mu$eV]} & theo &0.4 &5.05 & 5.45 \\   
                                            & exp & 1.55 & 4.00 & 5.55\footnote{From the measured total halflife $T_{1/2}$ of the excited state \cite{masuda}.}  \\                                       
    \hline
    \hline
    \end{tabular*}
    \caption{Reduced transition probabilities, branching ratios and radiative and total decay widths for the cross-band and in-band transitions obtained from theoretical predictions (theo), experimental data (exp) and a combination thereof. }
    \label{Used-param}
\end{table}
%%%%%%%%%%%%%%%%%%%%%%%%%%%%%%%%%%%%%%%

We note that the total decay rate of the upper state $|3\rangle$ obtained from theoretical estimates is very close to the experimental value from Ref.~\cite{masuda}. The branching ratios and especially the total branching ratio show the largest disagreement between theory and reconstruction from experimental values. At first sight, this discrepancy could stem from the different $B(M1)$ values appearing in the branching ratios. This holds true for the radiative branching ratio BR$_\gamma$, where the disagreement is however not that critical. However, the much larger and most relevant disagreement for the total branching ratio is caused by the already mentioned inconsistency in the $B(E2)$ values obtained from partial reconstruction based on experimental data.

\section{Numerical Results \label{numres}}

In this Section we present our numerical results for nuclear coherent population transfer in the considered $^{229}$Th three-level-$\Lambda$ system. We consider three different possible experimental scenarios and discuss their  feasibility. The first two  scenarios investigate the interaction of relativistic thorium ions accelerated in a storage ring considering (A)  UV radiation and strong acceleration or (B) x-ray radiation  from an XFEL source combined with moderate ion acceleration. 
The last scenario (C) addresses the interaction of a generic $^{229}$Th-doped solid sample and highly energetic x-rays from a cavity-based x-ray source.

\subsection{NCPT for highly accelerated nuclei}
\label{UV}
%%%%%%%%%%%%%%%%%%%%%%%%%%%%%%%%%%%%%%%%%%%%%%%%%%%%%%%%%%%%%%
The first scenario investigates the radiative coupling of ultra relativistic highly charged thorium ions with two UV laser fields.
The experimental implementation could be achieved at the GF,  which is an ambitious research tool for physics beyond colliders at CERN \cite{krasny2015gamma, budker2020, budker2021}.
The ultra relativistic acceleration of Th ions appears to be feasible given previous experiments at LHC with  highly ionized Pb ions, which have a similar mass to $^{229}$Th, where Lorentz factors up to $\gamma \approx2950$ were reached. We note that STIRAP in this scenario has been recently addressed in Refs.~\cite{budker2021,junlan22}.

For our purpose, we consider highly charged $^{229}$Th ions (or even bare nuclei) circulating at LHC  with $v \approx c$ and a Lorentz factor $\gamma = 2950$. We assume the ions are distributed in a bunch with kinetic energy fluctuations in the order of magnitude $\Delta \gamma / \gamma = \num{e-4}$.  
The transverse cross section of the ion beam can be approximated as $\pi r_\sigma^2$, where $r_\sigma = \SI{16}{\micro \metre}$ is the $1$-$\sigma$ radius of the ion bunch \cite{budker2021}.
The ion bunch has a revolution frequency of $f_b=\SI{11.2}{\kilo \hertz}$ and at the same time up to 1232 ion bunches can circulate in the storage ring. Each ion bunch carries up to \num{e8} $^{229}$Th ions \cite{budker2021}.
For resonant excitation we require photon energies $E_\lambda =\SI{29.19}{\kilo \electronvolt}/2\cdot2950 \approx \SI{4.95}{\electronvolt}$ for the pump field. The Stokes field photons are tuned such that a resonant coupling of isomer and second nuclear excited state occurs.
The initial pulse energy of each field is set to \SI{10}{\micro \joule}, which is then increased by a factor \num{e5} by means of cavity enhancement with a F\'abry-Perot interferometer to \SI{1}{\joule}. %Zitat.
The pulse duration is chosen to be \SI{3.7}{\pico \second} such that a broad spectral width is guaranteed.
The energy spread of the ion beam is accounted for approximatively by including a detuning in the calculation, which reads in the nuclear rest frame
\begin{equation}
    \Delta_P = 2\left(\gamma + \Delta \gamma \right)\omega_P -\omega_{\text{13}}  = \frac{\Delta \gamma }{\gamma} \cdot \omega_{13}\approx \Delta_S.
\end{equation}

The peak intensity of each laser field in the lab frame is defined as $I_0 = \frac{E_{\text{pul}}}{ t_{\text{pul}}\pi w^2 }$, where $w$ is the beam waist of the pulse. In addition, we assume the laser system has  a repetition rate $f_{\text{rep}}= 1232 \cdot f_b= \SI{13.8}{\mega \hertz} $ and it is perfectly synchronized to the circulating ion bunches.

We start by estimating the required intensities for NCPT via the adiabaticity and $\pi$-pulse criteria in Eqs.~\eqref{int-stirap} and \eqref{int-pi}. These estimates require knowledge of the nuclear reduced transition probabilities $B(M1)$ for the two transitions. Our intensity values for the experimentally deduced and theoretical $B(M1)$ sets in Table \ref{Used-param} are listed in 
 Table ~\ref{IntUV}. The ratio of the pump and Stokes intensities is related to the different reduced transition probabilities of the corresponding transitions, and is a factor $10$ for the experimental values and a factor $15$ for the theoretical values.

%%%%%%%%%%%%%%%%%%%%%%%%%%%%%%%%%%%%%%
\begin{table}[htpb!]
    \centering
    \begin{tabular}{lc|cc}
    \hline 
    \hline
  Method & $B(M1)$   & $I_{0,P}[\si{\watt \per \metre \squared}]$ & $I_{0,S}[\si{\watt \per \metre \squared}]$ \\
  \hline
STIRAP & exp & \num{2.75e24} & \num{2.77e23} \\
  & theo & \num{2.93e24}& \num{1.96e23} \\
  \hline
  $\pi$-pulses & exp & \num{8.64e22}& \num{8.70e21}  \\
  & theo & \num{9.22e22} & \num{6.16e21}\\
  \hline
  \hline
    \end{tabular}
    \caption{Laser intensities required for NCPT via STIRAP and $\pi$-pulses. The values were obtained  from the adiabaticity and $\pi-$pulse criteria for the given pulse duration $t_{\text{pul}}= \SI{3.7}{\pico \second}$ for the experimental (exp) and theoretical (theo) $B(M1)$ data sets. }
    \label{IntUV}
\end{table}
%%%%%%%%%%%%%%%%%%%%%%%%%%%%%%%%%%%%%%%%%%%%%%%%

We are interested in the fraction of nuclei which have reached the isomeric state after one sequence of pump and Stokes pulses.  Instead of the isomer population at the end of the laser pulses, we consider the transfer rate which also takes into account the ratio of ions unintentionally transferred to $\ket{3}$ and their subsequent decay after one ion revolution in the storage ring. The transfer rate for a single pulse sequence therefore reads
\begin{equation}
    \eta \approx \rho_{22} + \text{BR}_{32} \rho_{33}\left(1-\exp{\left(-\frac{\ln{2}}{\gamma  T_{1/2}f_b}\right)}\right).
\end{equation}
Here, $\rho_{ii}$  are the density matrix elements giving the fraction of population in state $\ket{i}$ and $\text{BR}_{32}$ the branching ratio of the in-band decay, respectively. The latter term takes into account the radiative decay of the intermediate state during one circulation in the storage ring. With aid of the experimental and theoretical reduced transition probabilitites $B(M1)$ we derive the radiative half life as $T_{1/2}^{\text{exp}}=\SI{25}{\nano \second}$ and $T_{1/2}^{\text{th}}=\SI{18}{\nano \second}$.

\begin{figure}
    \centering
    \includegraphics[width=8.5cm]{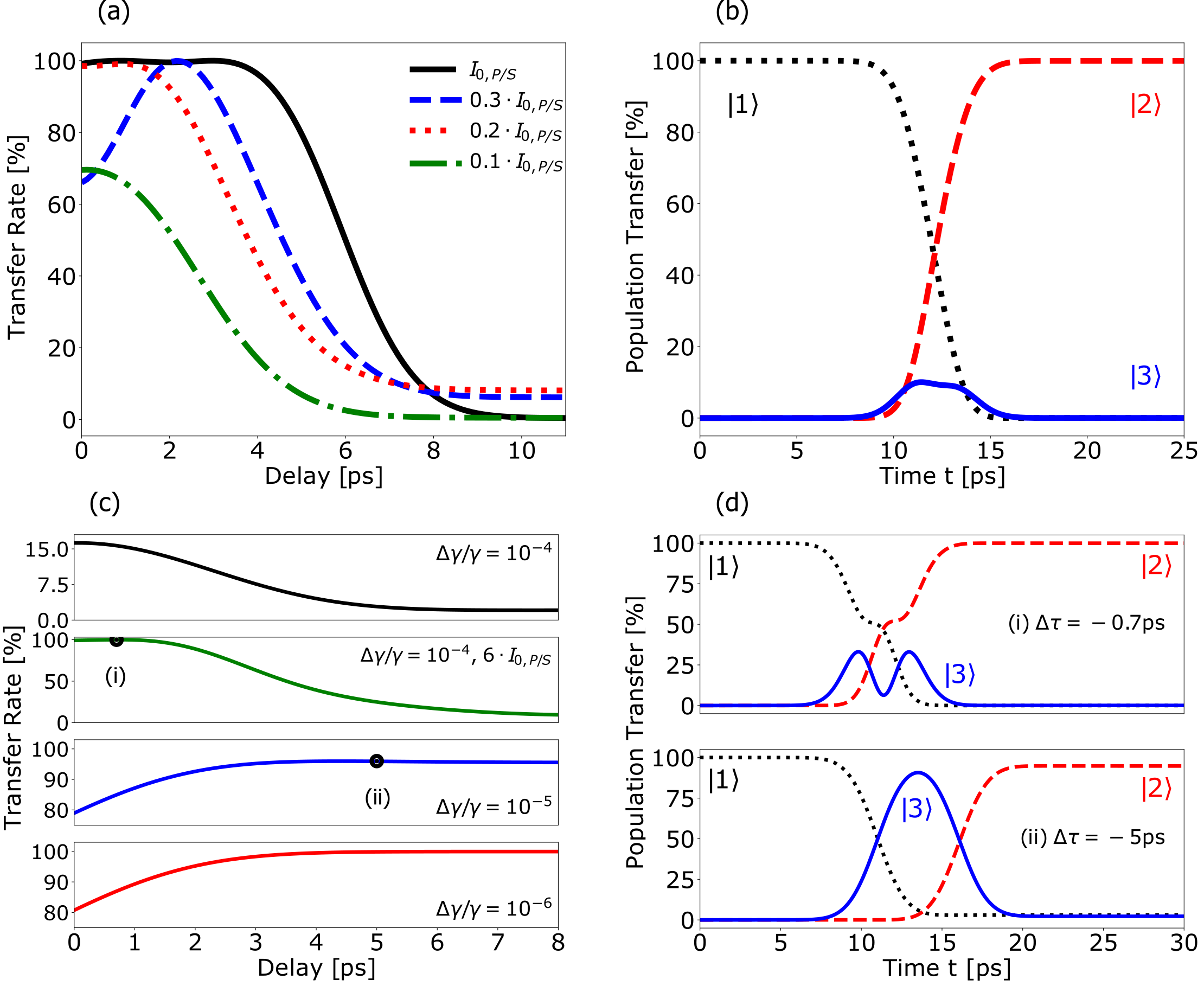}
    \caption{(a) STIRAP transfer rate as a function of time delay (in the lab frame) for different scaling factors of the laser intensities. (b) STIRAP population transfer as a function of time for $0.3\cdot I_{0,P/S}$ and $\Delta \tau = \SI{2.2}{\pico \second}$ (lab frame). (c) $\pi$-pulse transfer rate as a function of time delay between pulses $|\Delta \tau|$ (in the lab frame) for different ion energy spreads. (d) Single $\pi$-pulse population transfer sequence for (i) $\Delta \tau = \SI{-0.7}{\pico \second}$, $6\cdot I_{0,P/S}$ and $\Delta \gamma / \gamma = \num{e-4}$ and for (ii) $\Delta \tau =\SI{-5}{\pico \second}$, $I_{0,P/S}$ and $\Delta \gamma /\gamma = \num{e-5}$ (in the lab frame). }
    \label{uv}
\end{figure}

We start with NCPT via  STIRAP considering the intensities in Table \ref{IntUV} estimated for the experimental set of reduced transition probabilities $B(M1)$. Since the adiabaticity criterion is rather empirical, we consider also smaller intensities and vary the pulse delay to maximize the population transfer. 
To this end, the peak intensity of both Stokes and pump fields is scaled down in steps of 0.1.  Our numerical results  in Fig.~\ref{uv}(a) show that indeed, 100\% population transfer can be reached for a large range of delay times $\Delta\tau$ for the  intensity values $I_{0,P/S}$ of Table ~\ref{IntUV} (and, obviously, for any higher intensities). However, for particular values of $\Delta\tau$, STIRAP can be successful also for lower intensities. Figure \ref{uv}(a) presents the  transfer rate to the state $|2\rangle$ for scaled pump and Stokes intensities using three scaling factors 0.1, 0.2 and 0.3 where a trend change is observed. For $\Delta\tau=\SI{2.2}{\pico \second}$, 100\% population transfer can be achieved via STIRAP also for peak intensity values  0.3$I_{0,P/S}$. 
Our numerical results for the populations of the three states using these intensities are presented in Fig.~\ref{uv}(b).
Also the scaling 0.2 leads to transfer rates of almost unity. However, further lowering of the pulse intensities leads to incomplete population transfer rates.
Using the theoretical $B(M1)$ values  we obtain very similar results, and the transfer rates  deviate only by a few percent.

%%%%%%%%%%%%%%%%%%%%%%%%%%%%%%%%%%%%%
%%%%%%%%%%%%%%%%%%%%%%%%%%%%%%%%%%%%%

Next, we turn to NCPT via $\pi$-pulses. Our numerical results in Figure \ref{uv}(c)  show that transfer rates of unity cannot be reached due to the large detuning in the ensemble. Only about \SI{16}{\percent} of the population are transferred to the isomeric state for a small delay window. 
The transfer rates can be increased either by decreasing  the ion beam energy spread  or by increasing the laser intensity.
Indeed, as soon as we scale down the ion energy spread, the transfer rates significantly increase. For $\Delta \gamma / \gamma = \num{e-5}$ the transfer rates already increase to \SI{96}{\percent} and for $\Delta \gamma / \gamma = \num{e-6}$ transfer rates of unity are reached for a large time delay window. We note that a better ion beam quality would be beneficial also for NCPT via STIRAP.
Keeping $\Delta \gamma / \gamma = \num{e-4}$ fixed, transfer rates of approx. \SI{100}{\percent} can be reached by increasing the intensity. Considering the set of $B(M1)$ values deduced from experiments,  for the given pulse duration scaling by a factor 6 is required to obtain $\eta = \SI{100}{\percent}$ for $\Delta \tau = \SI{-.7}{\pico \second}$. This intensity corresponds to $\approx \SI{20}{\percent}$ of the intensity required for NCPT via STIRAP. Thereby, we observe that the sequence attains with the preceding pump pulse slightly higher rates, though for both cases very close to 100\%.

This scenario allows only a small delay window for transfer rates approaching unity, and a large pulse overlap is required. This is therefore less of the traditional $\pi$-pulse scheme but rather  a single train of coincident pulses, which builds a bridge between both NCPT schemes \cite{vitanov2012}.
This feature can be identified in the upper part of Fig.~\ref{uv}(d)
which illustrates a single population transfer event  for $\Delta \tau = \SI{-0.7}{\pico \second}$, $6\cdot I_{0,P/S}$ and $\Delta \gamma / \gamma = \num{e-4}$. In comparison a single event for $\Delta \tau =\SI{-5}{\pico \second}$, $I_{0,P/S}$ and $\Delta \gamma /\gamma = \num{e-5}$ is shown below. The difference is clearly visible. 
In  (i) only a small fraction of the population is pumped to the intermediate state compared to  (ii).

Let us apply our theoretical considerations to the  GF experimental setup and its given parameters.
Considering the  laser beam waists to be equal to the 1-$\sigma$ radius of the ion bunch $w_\sigma=\SI{16}{\micro \metre}$ yields intensities $I_{0,S}=I_{0,P}=\SI{3.36e20}{\watt \per \metre \squared}$ much smaller than the threshold deduced from the transfer criteria. Correspondingly, we can expect only low population transfer rates of much less than one percent.
The solution is to increase the intensity by stronger focusing and therefore smaller laser beam waist. 
The drawback is that less nuclei in the ion beam are irradiated by the lasers and only a fraction thereof can be promoted to the isomeric state. To investigate the trade-off between intensity and number of addressed nuclei we start by approximating the cross sections of the two laser beams  as circular and concentric, and in addition concentric to the ion beam cross section. Guided by the transfer criteria, we assume that the two Rabi frequencies for pump and Stokes pulses are equal for equal pulse durations. 
Considering the same laser energy for the two beams,
the smallest beam waist and therefore highest intensity is required for the pump laser, since the $B(M1)$ value is smaller for the $|1\rangle \rightarrow |3\rangle$ transition. The effective number of ions interacting with the laser photons is then $N_{\text{ion}}^{\text{eff}}= \left(\frac{w_{P}}{w_\sigma} \right)^2 N_{\text{ion}}$ with $w_P$  the beam waist of the pump  field. For a single NCPT process during one revolution of the ion beam, up to  $N_{\text{iso}}= N_{\text{ion}}^{\text{eff}}\, \eta$ isomers can be populated.

%%%%%%%%%%%%%%%%%%%%%%%%%%%%%%%%%%%%%%%%%%%%%%%%%
\begin{table*}[htpb!]
    \centering
    \begin{tabular*}{\linewidth}{@{\extracolsep{\fill}} l|cccccccc}
    \hline 
    \hline
  Method & $\Delta \gamma / \gamma $ & Scaling  & $w_P [\si{\micro \metre}]$  & $\Delta \tau [\si{\pico \second}]$  & $\eta [\si{\percent}]$ & $N_{\text{iso}}[\num{e4}]$   & $t_{ex}[\si{\second}] \gtrsim$ \\ 
  \hline
   STIRAP & \num{e-4} & 1 & 0.18 & 2.9 & 100 & 1.3 & 1.38   \\
  &\num{e-4}&0.3 & 0.32 & 2.2 & 100 & 4.0 &   0.42\\
  & \num{e-4}&0.2& 0.40 & 1 & 99.1 & 6.2 &   0.32\\
  &\num{e-4}&0.1& 0.56& 0.1 & 69.6 & 8.5   & 0.20\\
  \hline 
     $\pi$-pulses & \num{e-4} &  1 &  1 & -0.1 & 16.3 & 6.3  & 0.18\\
  &  \num{e-4}&6 & 0.41 & -0.7 & 100 & 6.5    & 0.29\\
 &  \num{e-5}&1 & 1 & -5 & 96 & 37.4    & 0.10\\
  & \num{e-6} &1 & 1 & -5 & 100 & 39.1    & 0.09\\
  \hline
  \hline
    \end{tabular*}
    \caption{Population transfer rate $\eta$,  number of produced isomers  $N_{\text{iso}}$ and time $t_{ex}$ required to approximately reach saturation  for the two NCPT schemes in scenario A.  The beam-waist of the Stokes field can be determined via $w_S=\sqrt{I_{0,P}/I_{0,S}} \cdot w_P$. For an equal focus ($w_P=w_S$) the pulse energy of the Stokes field should be lowered according to $E_S = I_{0,S}/I_{0,P}\cdot E_P $. See text for further explanations.}
    \label{performanceVUV}
\end{table*}
%%%%%%%%%%%%%%%%%%%%%%%%%%%%%%%%%%%%%%%%%%%%%%%%%

The results for both transfer schemes (for the experimental set of $B(M1)$ values) are summarized in Table~\ref{performanceVUV}. 
For STIRAP the number of produced isomers is rather small, since the intensities $I_{0,S}/I_{0,P}$ corresponds to rather small beam waists compared to the ion beam. Thus, only a small fraction of the ion beam is addressed and just a few ten thousand ions are promoted to the isomeric state. Scaling down the intensity values correspondingly leads to lower population transfer rates, but due to the larger beam size more isomers are excited.
A more systematic study reveals that choosing the same pump and Stokes beam waist (and therefore the same intensity for the two lasers) yields the maximum of approx. \num{1.5e5} isomers for $w_P=w_S =\SI{1.5}{\micro \metre}$ for delay times around $\Delta \tau = \SI{0.1}{\pico \second}$.  
In comparison, this focus also leads to better results than $w_P=w_S=r_\sigma=\SI{16}{\micro \metre}$, for which the number of populated isomers is only about \num{3.3e4}.
For NCPT with two $\pi$-pulses, our results show that the population transfer for $\Delta \gamma / \gamma =\num{e-4}$ does not deviate much from the STIRAP numbers. However, the implementation could be less challenging  due to the larger pump field waist. Provided it is possible to reduce the ion energy spread of the beam, then high transfer rates and consequently high isomer numbers can be produced as shown in the Table \ref{performanceVUV}.

For applications that make use of the produced isomers, it is very unlikely that a single pulse sequence is sufficient. We therefore investigate the scenario of repeated coherent pumping, considering that Stokes and pump lasers interact with the ion bunch with repetition rate $f_b$.
We assume that for each ion bunch revolution, the ions redistribute spatially in the beam. Thus, naively one could approximate that after the time $\frac{ N_{\text{ion}}}{ N_{\text{isomer}}} \cdot \frac{1}{f_b}$ (considered in the lab frame), a large fraction of ions in the storage ring are transferred to the isomeric state. However, this picture is not accurate since for increasing numbers of excited isomers, the laser pulses will also drive nuclear population from the isomeric state back to states $|1\rangle$ and $|3\rangle$ leading to  saturation. %Thus the value $t_{\text{ex}}$ should be regarded as a minimum value to obtain a large fraction of isomers in the beam.
%Thus the value $\frac{ N_{\text{ion}}}{ N_{\text{isomer}}} \cdot \frac{1}{f_b}$ should be regarded as a minimum value to obtain a large fraction of isomers in the beam

To model this scenario, we consider the same beam waist for both pump and Stokes pulses, such that both pulses address the same fraction of nuclei in the beam. 
To compensate in the intensity, we consider a smaller  Stokes pulse energy.
%Furthermore, we assume that the ions redistribute spatially in the beam during one revolution in the storage ring.
We model the spatial redistribution assuming that the ions with  nuclei in their respective states are distributed homogeneously within the bunch for each pulse sequence/ion bunch revolution. The  master equation is solved in a loop for the input parameters in Tab.~\ref{performanceVUV} where each iteration corresponds to a pulse sequence.
We obtain an isomer saturation of $\approx \SI{50}{\percent}$ for all NCPT scenarios with $\Delta \gamma / \gamma=\num{e-4}$ and $\approx \SI{70}{\percent}$ for all (non-overlapping) $\pi$-pulse scenarios with $\Delta \gamma / \gamma \leq \num{e-5}$.
Note that for a better ion energy spread ($\leq \num{e-5}$) also NCPT via STIRAP can reach a total isomer population of $\approx \SI{70}{\percent}$.
The excitation time $t_{ex}$ in Tab.~\ref{performanceVUV} corresponds to the approximate time which is required for the isomer population to reach the saturation limit.
We see by means of $t_{ex}$ that it is more advantageous for faster excitation to have a large beam waist. These times are for most scenarios approximately a factor 2 larger than the naive estimate $\frac{ N_{\text{ion}}}{ N_{\text{isomer}}} \cdot \frac{1}{f_b}$.

Incoherent pumping $\ket{1}\rightarrow \ket{3}$ with the pump laser only and the subsequent spontaneous decay to $\ket{2}$ provides an efficient alternative due to the advantageous branching ratio $\text{BR}_{32}$ and the slow radiative decay of the isomeric state. Thus, almost all ions in the bunch can reach the isomeric state. Our numerical simulations of a sequence of pump pulses leads to almost 100\% population of the isomeric state after a time which is approximately 5 times the $t_{ex}$ values in Table \ref{performanceVUV}.  However, it takes both STIRAP and incoherent pumping approximately the same time to reach the STIRAP saturation level of about 50\%.  
Thus, the advantage of STIRAP  to reduce the excitation time (albeit with a more complicated experimental setup) only comes into play with large if sufficient intensities are available for laser beams with waists which approach the ion beam radius.

\subsection{NCPT with moderately accelerated nuclei}
\label{xfelTransfer}
%%%%%%%%%%%%%%%%%%%%%%%%%%%%%%%%%%%%%%%%%%%%%%%%%%%%%%%%%%%%
The second excitation scheme considers the radiative coupling of moderately accelerated thorium ions with coherent x-ray pulses from an  XFEL. In comparison to optical laser systems,  XFEL pulses may lack temporal coherence  and most currently operating facilities only have repetition rates of a few tens up to a hundred Hz with a few exceptions \cite{HUANG2021}.
The photon energies of XFELs range from hundred eV (soft x-rays) up to \SI{25}{\kilo \electronvolt} \cite{desy_overview}.
Unfortunately, these energies are too low to directly excite the 29 keV level in $^{229}$Th, so some acceleration of the nuclear beam is required. For our purposes, we consider the pump field  operating at a photon energy of \SI{3.5}{\kilo \electronvolt}, while the Stokes field is operating at a slightly smaller energy.
For these energies, the ions have to be accelerated to $\gamma \approx 4.2$ for resonance. This small Lorentz factor is not available at CERN since there $\gamma_{\text{min}}=20$ \cite{budker2021}. Thus, the experiment would require a smaller storage ring. For instance, the high energy storage ring (HESR) at GSI can deliver sufficiently small  Lorentz factors for heavy ions \cite{wiley_GSI}. Compared to LHC ($\approx \SI{40}{\hour}$)\cite{budker2021}, the lifetime of U$^{90+}$ (similar mass as  Th) with $\gamma \approx 4$ is expected to be $\approx\SI{3000}{\second}$ \cite{GSI_lifetime}.

%The discussed setup is rather challenging to implement experimentally due to the infrastructure. An experimenter requires the simultaneous availability of a storage ring and two XFEL sources either from a suitable two color operation or two XFEL beamlines. However, up to this point, there are no tabletop solutions for this purpose and rather small facilities already have lengths of \SI{0.7}{\kilo \metre} \cite{swissXFEL}.

We assume the ion bunch carries up to \num{e8} ions with a transverse 1-$\sigma$ radius of \SI{2.1}{\milli \metre} and the circulation frequency of an ion bunch is $f_b = \SI{522}{\kilo \hertz}$. The large ion beam size is correlated with the relativistic Lorentz factor $\gamma$ since the adiabatic damping requires larger particle momentum \cite{acceleratorPhys}.
If not mentioned otherwise, we consider also for this case the ion energy spread  $\Delta \gamma / \gamma =\num{e-4}$. Again, this then leads to a rest frame detuning $\Delta_P = \Delta_S \approx \Delta \gamma / \gamma \cdot \omega_{\text{13}}$ in the ensemble.
The generic laser parameters we consider \cite{xfel_overview} are   pulse energy  \SI{1}{\milli \joule},  pulse duration \SI{30}{\femto \second} and  repetition rate $f_{\text{rep}}=\SI{60}{\hertz}$.
We assume the pulses have full temporal coherence, which is not far from what has been reported in Ref.~\cite{nam2019}. 
Since the repetition rate of the XFEL is small compared to the circulation frequency, an ion bunch can circulate (in the HESR case) $f_b/f_{\text{rep}}=8700$ times before another XFEL pulse sequence arrives. Therefore, the transfer rate after one pulse sequence can be approximated as $\eta \approx \rho_{22}+\text{BR}_{32}\rho_{33}$.
 The required theoretical and experimental intensities for NCPT via STIRAP and $\pi$-pulses are listed in Tab.~\ref{performanceX}. Compared to the figures in Table~\ref{IntUV}, it is clear that this setup  requires much larger laser intensities to be successful due to the fs XFEL pulse durations.

%%%%%%%%%%%%%%%%%%%%%%%%%%%%%%%%%%%
\begin{table}[htpb!]
    \centering
    \begin{tabular}{lc|cc}
    \hline 
    \hline
  Method & $B(M1)$   & $I_{0,P}[\si{\watt \per \metre \squared}]$ & $I_{0,S}[\si{\watt \per \metre \squared}]$ \\
  \hline
STIRAP & exp & \num{4.18e28} & \num{4.21e27} \\
  & theo & \num{4.46e28}& \num{2.98e27} \\
  \hline
  $\pi$-pulses & exp & \num{1.31e27}& \num{1.32e26}  \\
  & theo & \num{1.40e27} & \num{9.38e25}\\
  \hline
  \hline
    \end{tabular}
    \caption{Laser intensities required for NCPT via STIRAP and $\pi$-pulses for scenario B and a pulse duration of $\approx \SI{30}{\femto \second}$. See caption of Table~\ref{IntUV} for further explanations.}
    \label{performanceX}
\end{table}
%%%%%%%%%%%%%%%%%%%%%%%%%%%%%%%%%%%

\begin{figure}[htb!]
    \centering
    \includegraphics[width=9cm]{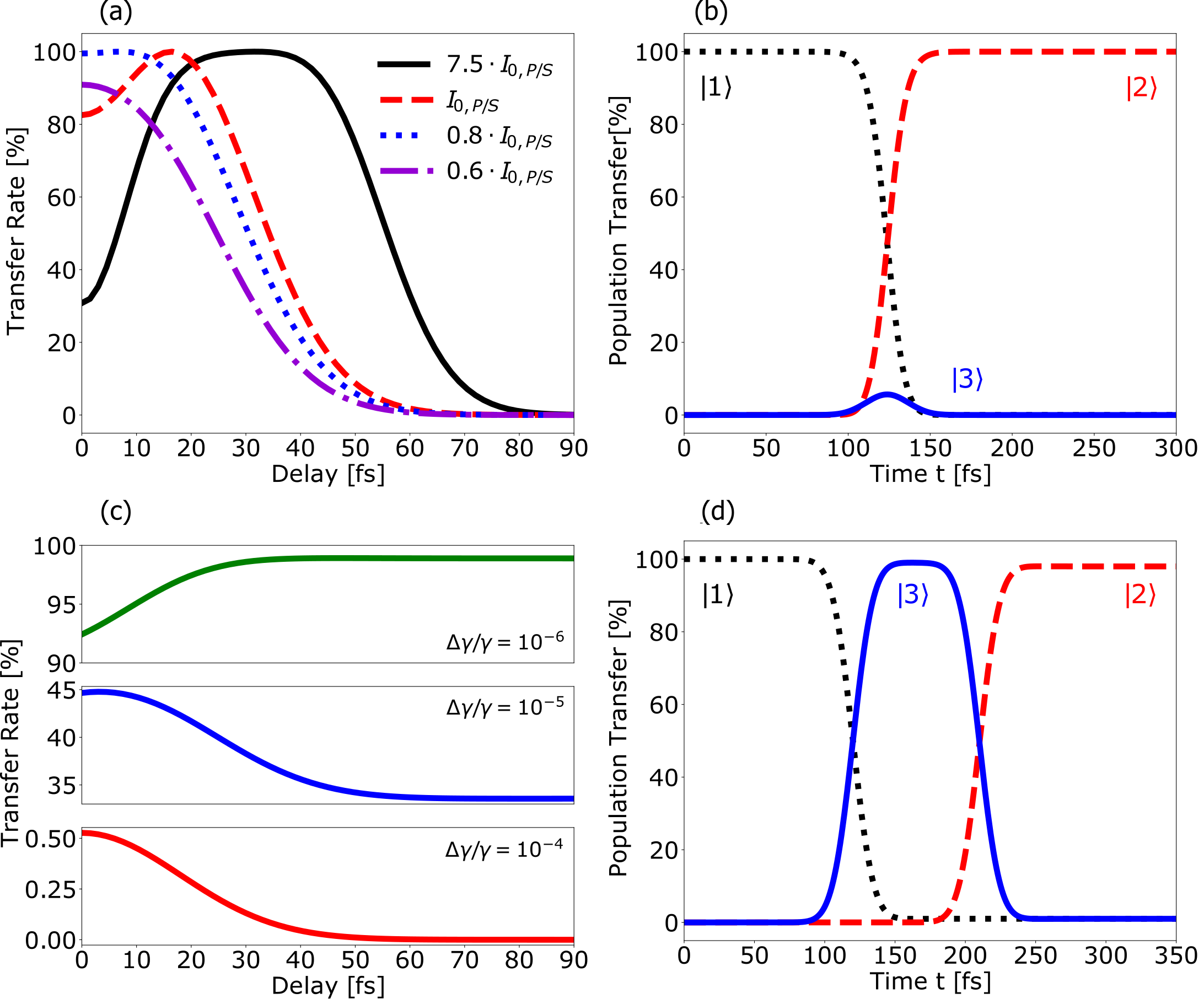}
    \caption{(a) STIRAP transfer rate as a function of delay (in the lab frame) for different XFEL intensity values  for $\Delta \gamma / \gamma = \num{e-4}$. (b) Single STIRAP sequence (lab frame) for $0.8 \cdot I_{0,P/S}$ and $\Delta \tau = \SI{7.5}{\femto \second}$. (c) $\pi$-pulse transfer rate as a function of delay $|\Delta \tau|$ (lab frame) for different ion energy spreads. (d) Single $\pi$-pulse sequence (lab frame) for $\Delta \gamma / \gamma = \num{e-6}$ and $\Delta \tau = \SI{-90}{\femto \second}$. }
    \label{XFEL_Paper}
\end{figure}

In the following we present our results for the case of the $B(M1)$ values obtained from experimental data. 
 Following the discussion in  Sec.~\ref{UV}, at first NCPT via STIRAP is discussed. Starting from the intensities delivered by the adiabaticity criterion in Table~\ref{performanceX}, we investigate in Fig.~\ref{XFEL_Paper}(a) the transfer rate as a function of delay time between XFEL pulses for down-scaled intensity values. 
Transfer rates of unity are reached until $0.8 \cdot I_{0,P/S}$. The population transfer after a single STIRAP sequence for this case is shown in Fig.~\ref{XFEL_Paper}(b). From a scaling of 0.7 the transfer rate slightly decreases to $\approx \SI{98.7}{\percent}$, while a scaling of 0.6 leads to a drop of $\eta$ to $\approx \SI{91}{\percent}$. In contrast to the results in Sec.~\ref{UV}, the broader plateaus of the transfer rate as a function of delay time appear not for the reference intensity $I_{0,P/S}$, but for the down-scaled (and also for upscaled) values. Thus, also smaller intensities allow for a more feasible and robust STIRAP implementation.

Figure \ref{XFEL_Paper}(c) presents our numerical results of the transfer rate for  a $\pi$-pulse sequence. Once more, we observe a rather bad performance due to the detuning caused by the ion beam energy spread. Only about \SI{.5}{\percent} are transferred to the isomeric state. This is related to the very short pulse durations of the lasers and the resulting broad spectral width. Larger transfer rates can be obtained either by increasing the XFEL intensity, or by decreasing the ion energy spread.
Once the intensity is scaled up  by a factor of approx. 25 we again reach transfer rates approaching unity for large pulse overlaps.
In turn, keeping the intensity set to $I_{0,P/S}$, for
 $\Delta \gamma / \gamma= \num{e-6}$ transfer rates of approx. \SI{99}{\percent} can be reached with high stability and the characteristic plateau for large delays. In the case of $\Delta \gamma /\gamma = \num{e-5}$, transfer rates of unity are only achievable  with a up-scaling of the intensities and within a short time interval around \SI{15}{\pico \second} with a large pulse overlap.

%%%%%%%%%%%%%%%%%%%%%%%%%%%%%%%%%%%%%%%
%%%%%%%%%%%%%%%%%%%%%%%%%%%%%%%%%%%%%%%%

Our simulations show that scenario B is problematic from several points of view. The large intensities required for NCPT imply either a very high and so far unavailable  pulse energy, or a strong focus. For instance, for intensity values of $0.8 \cdot I_{0,P/S}$ and pulse energy 1 mJ, the required beam waists are  $w_p=\SI{0.56}{\nano \metre}$ and $w_s=\SI{1.77}{\nano \metre}$, exceeding the typical focusing limits of a few nm \cite{xrayfoc2009,xrayfoc2013,xrayfoc20}. With stronger focus the number of addressed isomers is also diminishing. In this example, the effective number of ions significantly decreases to $\approx \num{7e-6}$, which means hardly any thorium ions are addressed.
Even using an unrealistic pulse energy input for the calculation, the isomer population does not approach the same levels as scenario A due to the much lower pulse repetition rates. Finally, setup B is logistically challenging because it requires both an XFEL and an ion accelerator at the same facility. We conclude that scenario B is less likely to be experimentally implemented.

\subsection{NCPT with $^{229}$Th nuclei at rest}
%%%%%%%%%%%%%%%%%%%%%%%%%%%%%%%%%%%%%%%%%%%%%%%%%%%%%%%%%
The last scenario investigates the case of a fixed thorium sample irradiated by resonant and  fully coherent x-ray pulses.  These pulses could be delivered by a next generation cavity-based FEL lasing source, the so-called XFEL oscillator (XFELO) \cite{XFELO_Prop, xfeloAdams,qin, lindberg2011}.
This lasing source is expected to provide radiation with large temporal coherence times and photon energies up to \SI{25}{\kilo \electronvolt} in basic operation. Theoretical studies have even shown that an XFELO can generate photons with energies up to \SI{60}{\kilo \electronvolt} through high harmonic generation \cite{qin_2018}. In the following we consider the XFELO parameters for basic operation despite requiring a photon energy of  $\SI{29.19}{\kilo \electronvolt}$.

X-ray pumping of the 29 keV level in  $^{229}$Th using nuclear resonant scattering synchrotron radiation and a fixed target has been reported in Ref.~\cite{masuda}. The target comprised of $^{229}$Th-doped thorium oxide sealed in Be cover plates. One could imagine also using a VUV-transparent target such as $^{229}$Th-doped CaF$_2$, which has an experimentally measured band gap of $\approx \SI{12}{\electronvolt}$ \cite{bandgapCaF2, caf2}. This would have the advantage that IC is not allowed from the isomeric state and in addition,  due to transparency, VUV photons from the radiative decay of the isomer can be detected. It is expected that both types of samples would experience damage from the highly energetic and intense radiation, such that a tape station for the target shifting the impact point after each pulse sequence would be required. 

Since the $^{229}$Th atoms or ions in the sample are not highly charged, the IC decay channel of the 29 keV state is energetically permitted. Thus, all nuclei which are pumped to the second excited state will decay via IC to both isomer and ground states. The total half-life of the 29 keV state derived from theory using the parameters $q_s=g_R = 0.6$ is $T_{1/2}=\SI{84}{\pico \second}$,  in good agreement with the experimental value reported in Ref.~\cite{masuda}. In the following we use the theoretical $B(M1)$ and branching ratio values for the calculation. In addition, we neglect detuning and set  $\Delta_P = \Delta_S = 0$. The transfer rate is given by  $\eta \approx  \rho_{22}+\text{BR}_{32}\rho_{33}$, where we use $\text{BR}_{32}=\SI{92.5}{\percent}$.

%%%%%%%%%%%%%%%%%%%%%%%%%%%%%%%%%%%%%%%%%%%%%%%%%%%%
\begin{table*}[htpb!]
    \centering
    \begin{tabular*}{\linewidth}{@{\extracolsep{\fill}} lcc|ccc|cccc}
    \hline 
    \hline
  Set  & $t_{\text{pul}}[\si{\pico \second}]$  & $E_{\text{pul}}[\si{\micro \joule}]$  & Method  & $I_{0,P}[\si{\watt \per \metre \squared}]$ & $I_{0,S}[\si{\watt \per \metre \squared}]$  & Scaling & $\Delta \tau[\si{\pico \second}]$ & $\eta [\si{\percent}]$ & $w_{P} [\si{\nano \metre}]$ \\ 
  \hline
   \cite{qin} & 0.7& 28 & STIRAP & \num{8.20e25} & \num{5.48e24} & 3-0.04  &  0.7-0.01&  100-97.7 & 0.23-1.97\\
  & 0.7 & 28 & $\pi$-pulses & \num{2.58e24} & \num{1.72e23} & 1 & -1.15  & 99.8  & 2.2 \\
  \hline 
       \cite{XfeloPetra} & 12 & 5 & STIRAP & \num{2.79e23} & \num{1.87e22} & 3-0.1 & 12.0-4.0 & 100-98.7  & 0.4-2.2 \\
  & 12 & 5 &  $\pi$-pulses & \num{8.76e21} & \num{5.86e20} & 1&  -16  & 98.2  & 3.9  \\
  \hline
  \hline
    \end{tabular*}
    \caption{XFELO pulse duration   $t_{\text{pul}}$ and  energy  $E_{\text{pul}}$ for two parameter sets discussed in the literature (left). 
    The corresponding laser intensities required for NCPT via STIRAP and $\pi$-pulses for theoretical $B(M1)$ values are given in the middle column split. The calculated ranges of  NCPT transfer rates $\eta$ are presented in the right column split. In vicinity of the adiabatic criterion intensity value, $\Delta \tau = t_{\text{pul}}$ corresponds to the optimum delay, while for shrinking intensity the maximum shifts towards smaller delays.
    The beam waist of the Stokes field can be determined via $w_S=\sqrt{I_{0,P}/I_{0,S}} \cdot w_P$.}
    \label{performance}
\end{table*}
%%%%%%%%%%%%%%%%%%%%%%%%%%%%%%%%%%%%%%%%%%%%%%%%%%%%%%

%%%%%%%%%%%%%%%%%%%%%%%%%%%%%%%%%%%%%%%%%%%%%%%%%%%%%%
\begin{figure}[htpb!]
    \centering
    \includegraphics[width=8.5cm]{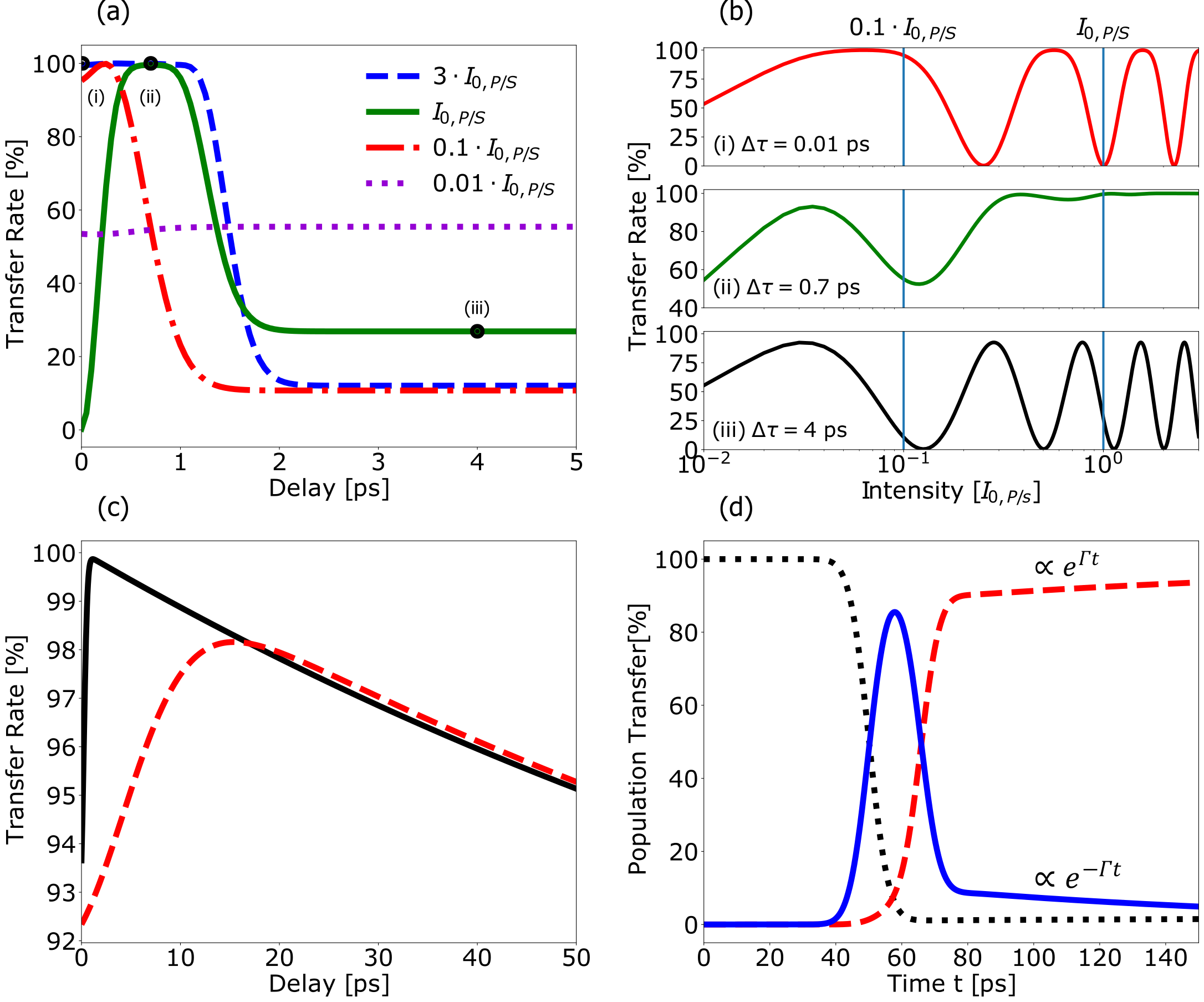}
    \caption{(a) STIRAP transfer rate as a function of pulse delay with scaled intensities (XFELO parameters from \cite{qin}). (b) Related transfer rate as a function of intensity for different pulse delays.  The vertical lines on the right and left-hand side correspond to $3\cdot I_{0,P/S}$ and $0.01 \cdot I_{0,P/S}$.(c) $\pi$-pulse transfer rate as a function of pulse delay $|\Delta \tau|$ for XFELO parameters  \cite{qin} (solid line), and  \cite{XfeloPetra} (dashed line). (d) Single $\pi$-pulse sequence for set \cite{XfeloPetra} with delay $\Delta \tau = \SI{-16}{\pico \second}$.}
    \label{Xfelo}
\end{figure}
%%%%%%%%%%%%%%%%%%%%%%%%%%%%%%%%%%%%%%%%%%%%%%%%%%%%%%

We proceed once more to calculate the x-ray pulse intensities required by the adiabatic and $\pi$-pulse criteria. The results are listed  in Table \ref{performance} for two sets of XFELO parameters which mainly differ in their x-ray pulse duration and pulse energy. We first address NCPT via STIRAP and investigate the  population transfer behaviour for different intensity scaling starting from the generic values  $I_{0,P/S}$ in Table \ref{performance}. 
 The transfer rate as a function of delay for the parameter set of Ref.~\cite{qin} is presented in Fig.~\ref{Xfelo}(a). For a large pulse intensity of  $3 \cdot I_{0,P/S}$, STIRAP appears to be robust and 100\% transfer rates  are reached for  a plateau of time delay values on the scale of the pulse duration. Once the intensity is decreased towards $I_{0,P/S}$ and less, the plateau narrows visibly and the peak of 100\% NCPT shifts towards smaller pulse delays. Transfer rates of almost unity are still reached for intensities of as low as $0.04 \cdot I_{0,P/S}$ ($\eta = \SI{97.7}{\percent}$). 
 
 The overall behaviour presented in Fig.~\ref{Xfelo}(a) displays some interesting features. 
 For instance at $\Delta \tau = \SI{0.01}{\pico \second}$ and an intensity  scaling of 1, the transfer rate is almost vanishing. Moreover, the transfer rates for large time delays between pulses saturate at different values for different intensities.  To understand better this behaviour, the transfer rate as a function of intensity $\eta \left(I \right)$ is calculated for three different pulse delays $\Delta\tau$ marked by bullet points in Fig.~\ref{Xfelo}(a). The numerical results are shown in  Fig.~\ref{Xfelo}(b). For small (i) and large (iii) delay between pulses, $\eta \left(I \right)$ shows an oscillatory behaviour. However, due to the large pulse overlap, transfer rates of unity can be reached for small delays, while for largely delayed pulses transfer rates of only $\SI{92.5}{\percent}$ are reached. That is because the Stokes pulse does not affect the system in this case.
Depending on the pump pulse intensity, the population is either  completely pumped to $\ket{3}$ ($\pi$-pulse) or stays in the ground state ($2 \pi$-pulse) or it is only partially pumped. 
In case of a single $\pi$-pulse configuration the population is only pumped to the intermediate state $\ket{3}$ from which it  subsequently decays. Due to the branching ratio of the in-band transition, only $\SI{92.5}{\percent}$ of the nuclei can be promoted to the isomeric state.

In case of an optimal robust STIRAP delay $(\Delta \tau \approx t_{\text{pul}})$ (ii), $\eta \left(I \right)$ shows the oscillatory behaviour only for small intensities. Once $I_{0,P/S}$ is exceeded, the oscillation stops and the transfer rate saturates at $\approx \SI{100}{\percent}$ which then corresponds to robust STIRAP.
We note that the numerical results for the XFELO parameters from Ref.~\cite{XfeloPetra} show a similar behaviour. For this case,  the transfer rates drop below $\SI{99}{\percent}$ at approximately $0.1 \cdot I_{0,P/S}$.
We summarize the most important results for both XFELO parameter sets in Tab.~\ref{performance}. 

We now turn to the case of  NCPT via two subsequent $\pi$-pules. Figure \ref{Xfelo}(c) shows the transfer rate  $\eta \left( \tau \right)$ for both XFELO parameter sets. For both cases, the transfer rates can be large, and decrease for large  delay times.
The reason for this is the short half-life of the intermediate state due to the open IC channel of state $\ket{3}$. This decay channel affects the population transfer already on ps-time scale.  Therefore, the total population in the isomeric state decreases due to losses from $\ket{3}\rightarrow \ket{1}$ and re-pumping from $\ket{2}\rightarrow \ket{3}$ and  subsequent decay. 
This effect is also the reason why the set \cite{qin} achieves slightly higher transfer rates, since  it operates on a shorter timescale compared to set \cite{XfeloPetra}.
In Fig.~\ref{Xfelo}(d) one can observe the exponential decay due to IC  during the population transfer. In this sequence the small slope in the population of $\ket{2}$ and $\ket{3}$ corresponds to an exponential decay/gain since for small times the exponential factor approaches to $e^{\pm \Gamma t} \approx 1 \pm \Gamma t$.

We note here that also for this scenario, the required laser intensities are very large and correspond to very small focal spots.
For instance, the pump pulse requires a beam waist of $\SI{2.2}{\nano \metre}$ ($\SI{3.9}{\nano \metre}$) for the parameters in Ref.~\cite{qin} (Ref.~\cite{XfeloPetra}) to fulfill the $\pi$-pulse criterion. To avoid different laser beam waists we once more can choose a lower pulse energy for the Stokes field. The very tight focusing also means that just a small volume in the solid-state sample is addressed by the laser pulses. However, the large dopant density of the sample leads to a number of irradiated nuclei comparable to the ion beam case.
Considering the sample used in Ref.~\cite{masuda} (thickness $D=\SI{0.2}{\milli \metre}$, diameter $L=\SI{0.4}{\milli \metre}$, dopant number $N=\num{6.3e14}$)  the number of irradiated nuclei can be estimated under the assumption of a disk-shaped 
sample and a homogeneous dopant distribution as $N_{\text{irradiated}}=N \frac{ w^2}{L^2}$. For a beam waist of $\SI{2.2}{\nano \metre}$ ($\SI{3.9}{\nano \metre}$) one can expect to irradiate up to \num{1.9e4} (\num{6e4}) thorium atoms.

\section{Conclusions  \label{concl}}
The second nuclear excited state of $^{229}$Th opens the possibility to indirectly reach the isomeric state at 8 eV via x-ray pumping. We have investigated this possibility following two directions of study. First, we have focused on the nuclear transition rates which characterize the first three levels of the $^{229}$Th nucleus. These values have been deduced within a nuclear structure model, and discussed in the context of the few available experimental data values. Our analysis points out at inconsistencies between the central experimental values deduced from experiments, allowing to identify the error bar interval which can accommodate all measured data.
Our findings summarized in Table \ref{Used-param} will be useful for future experiments employing x-ray pumping.

Second, we have studied NCPT for isomer population in  x-ray quantum optics schemes involving STIRAP and $\pi$-pulses configurations. These schemes are experimentally challenging and typically require large pulse intensities. From the three possible setups addressed, our simulations have identified the GF scenario with two UV laser beams interacting with relativistically accelerated $^{229}$Th ions as the most promising one. 
Our results show that NCPT with two subsequent $\pi$-pulses could be implemented with less experimental effort since the required intensities are smaller than in the STIRAP case.
However, the $\pi$-pulse excitation scheme necessitates almost full resonance which is usually not provided in storage ring experiments due to the ion beam energy spread.
Thus, STIRAP yields a robust alternative for coherent excitation although a detuning is present.
%Thereby, we have seen that the implementation of NCPT with $\pi$-pulses requires less experimental effort due to the smaller intensities which are required. However, this excitation scheme requires almost full resonance which is usually not provided in storage ring experiments. Therefore, STIRAP yields a robust alternative for coherent excitation although a detuning is present.
%\textcolor{blue}{Here some conclusion on the comparison between STIRAP and Pi-pulses should be mentioned.}
Due to an advantageous in-band branching ratio, direct x-ray pumping to the 29.19 keV state might prove to be competitive, provided strong XFEL pulses can be used at this energy.

\begin{acknowledgments}
%%%%%%%%%%%%%%%%%%%%%%%%%%%%%%%%%%%%%%%%%%%%%%%%%%%%%%%%%%%%%%%%%%%%%%%%%%
We thank Brenden S. Nickerson for fruitful discussions. 
AP gratefully acknowledges funding from the DFG in the framework of the Heisenberg Program.
This work is part of the ThoriumNuclearClock project that has received funding from the European Research Council (ERC) under the European Union’s Horizon 2020 research and innovation programme (Grant agreement No. 856415).
NM gratefully acknowledges the support by the Bulgarian National Science Fund (BNSF) under
contract No. KP-06-N48/1.

\end{acknowledgments}

\bibliography{refs}

\end{document}